\documentclass[format=acmsmall, review=false, screen=true]{acmart}

\usepackage{booktabs}
\usepackage{tikz}
\usetikzlibrary{external}
\usepackage{textcomp}
\usepackage[utf8]{inputenc}
\usepackage{enumitem}

\usepackage{makecell}
\usepackage{threeparttable}
\usepackage{longtable}
\usepackage{multirow}
\usepackage{tabularx}
\usepackage{tabulary}
\usepackage{array}
\newcommand{\PreserveBackslash}[1]{\let\temp=\\#1\let\\=\temp}

\newcolumntype{C}[1]{>{\PreserveBackslash\centering}p{#1}}
\newcolumntype{R}[1]{>{\PreserveBackslash\raggedleft}p{#1}}
\newcolumntype{L}[1]{>{\PreserveBackslash\raggedright}p{#1}}
\usepackage{amssymb}
\usepackage{eqnarray}
\usepackage{amsmath}
\usepackage{mathtools}
\usepackage{epstopdf}
\usepackage{graphicx}
\usepackage{subfig}
\usepackage{caption}
\usepackage{color}
\usepackage{url}
\usepackage{hyperref}
\usepackage{bm}
\usepackage[linesnumbered,ruled]{algorithm2e}

\setcopyright{acmlicensed}

\acmDOI{0000001.0000001}

\acmJournal{TOIS}
\acmVolume{0}
\acmNumber{0}
\acmArticle{0}
\acmYear{0000}
\acmMonth{0}
\copyrightyear{0000}

\begin{document}
\title{Attentive Long Short-Term Preference Modeling for Personalized Product Search}
\author{Yangyang Guo}
\orcid{}
\affiliation{%
  \institution{Shandong University}
  \country{China}
}
\email{guoyang.eric@gmail.com}

\author{Zhiyong Cheng$^{\dag}$}
\affiliation{%
  \institution{ Qilu University of Technology (Shandong Academy of Sciences)}
  \streetaddress{ Shandong Computer Science Center (National Supercomputer Center in Jinan), Shandong Artificial Intelligence Institute}
  \country{China}
 }
\email{jason.zy.cheng@gmail.com}

\author{Liqiang Nie$^{\dag}$}
\affiliation{%
  \institution{Shandong University}
  \country{China}
}
\email{nieliqiang@gmail.com}

\author{Yinglong Wang}
\affiliation{%
  \institution{Qilu University of Technology (Shandong Academy of Sciences)}
  \country{China}
}
\email{wangyl@sdas.org}

\author{Jun Ma}
\affiliation{%
  \institution{Shandong University}
  \country{China}
}
\email{majun@sdu.edu.cn}

\author{Mohan Kankanhalli}
\affiliation{%
  \institution{National University of Singapore}
  \country{Singapore}
}
\email{mohan@comp.nus.edu.sg}

\thanks{$^{\dag}$ Corresponding Author.\\ This work was finished when Yangyang Guo was a visiting student at the National University of Singapore. The first author claims that this work is under the supervision of Dr. Zhiyong Cheng and Dr. Liqiang Nie.}

\begin{abstract}
E-commerce users may expect different products even for the same query, due to their diverse personal preferences. It is well-known that there are two types of preferences:  long-term ones and short-term ones. The former refers to users' inherent purchasing bias and evolves slowly. By contrast, the latter reflects users' purchasing inclination in a relatively short period. They both affect users' current purchasing intentions. However, few research efforts have been dedicated to jointly model them for the personalized product search. To this end, we propose a novel Attentive Long Short-Term Preference model, dubbed as ALSTP, for personalized product search. Our model adopts the neural networks approach to learn and integrate the long- and short-term user preferences with the current query for the personalized product search. In particular, two attention networks are designed to distinguish which factors in the short-term as well as long-term user preferences are more relevant to the current query. This unique design enables our model to capture users' current search intentions more accurately. Our work is the first to apply attention mechanisms to integrate both long- and short-term user preferences with the given query for the personalized search. Extensive experiments over four Amazon product datasets show that our model significantly outperforms several state-of-the-art product search methods in terms of different evaluation metrics.

\end{abstract}



\maketitle

\section{Introduction}\label{introduction}
Nowadays, e-commerce has become very popular with the flourishing of the Internet. Its convenient access to an enormous variety of goods enables people to shop almost all products at home. When users on e-commerce websites like Amazon\footnote{\href{www.amazon.com}{https://www.amazon.com.}}  intend to purchase a product, they usually pick their desired ones among millions of items by searching. The standard scenario for the online product search is that a user submits a query, and then the search engine returns a ranked list of products relevant to the given query. Typically, queries submitted by users comprise only a few keywords (e.g., \emph{white T-shirt for men}), which are usually too short or ambiguous to express the users' needs precisely, resulting in unsatisfactory search results. Besides, users' preferences on products could be very diverse (due to different backgrounds, i.e., \emph{age}, \emph{gender}, \emph{income}) or strongly affected by the current contexts (e.g., \emph{season}, \emph{location}). Thereby it is not appropriate to return the same search results to different users for the same query. In the light of this, considering the user's personal intentions under the current contexts and aiming to return relevant products to the given query, the so-called personalized product search, plays a pivotal role in meeting the user's current shopping needs.

The prerequisite for a good personalized product search engine is to accurately model the user preferences and effectively integrate them with the current query. It is well-recognized that there are two types of user preferences~\cite{xiang2010temporal, bennett2012modeling}: long-term ones and short-term ones. The former refers to the user's inherent and relatively stable (evolving slowly) purchasing bias, such as \emph{favorite colors}, \emph{preferred fashion designs}, \emph{fitting sizes} and \emph{consumption level}, which is imperceptibly influenced by the user's personal backgrounds, like \emph{age}, \emph{upbringing}, \emph{marriage}, \emph{education} and \emph{income}. By contrast, the short-term user preference conveys user's purchasing intention in a relatively short period. It is affected by incidentally transient events, such as \emph{new product release}, \emph{season change} and \emph{special personal occasions} like \emph{birthday} ~\cite{xiang2010temporal}, which can be inferred from the user's recently purchased products. Compared to the long-term user preference, the short-term one changes more frequently and drastically.

Traditional approaches to product search~\cite{duan2015mining, duan2013probabilistic, duan2013supporting, van2016learning}  often employ simple matching between the queries and products without harnessing the user's specific attributes. They hardly characterize the user specificity, let alone the heterogeneity. It thus often leads to sub-optimal performance due to the user's diverse expectations. In view of this, personalization becomes quite necessary for the product search. In fact, personalized search has been widely studied in literature over the past few years. Thereinto, the study of personalized web search incorporating the user preference is the most related sub-direction to our work, and it can be roughly divided into two categories: short-term session-based\footnote{A session comprises a set of previous interactions containing past submitted queries and clicked records within a specific time limit~\cite{kacem2017emphasizing}.}  and long-term profile-based web search. Approaches in the first category ~\cite{kacem2017emphasizing, sriram2004session, daoud2009session} capture the short-term user preference from search sessions. Nevertheless, different from the web search whereby a session often contains plenty of queries and rich click results, the session data in the product search~\footnote{Pairs of queries and the corresponding purchased products.} are usually too sparse to train a good personalized search model. Methods in the second category ~\cite{matthijs2011personalizing, tan2006mining} model the long-term user preference based on user's complete browsing behaviors, yet they suffer from two limitations. First, they do not fully exploit user's recent behaviors, containing important contextual information for the current search intention that could reflect the user's recent preference (the short-term preference), e.g., a recently bought \emph{computer} is often followed by accessories like \emph{a mouse} or \emph{a keyboard}. Second, they assume that the long-term user preference is stable yet in fact, it changes over time slightly and slowly~\cite{white2010predicting, ustinovskiy2013personalization}. Recently, Ai et al. ~\cite{ai2017learning} have presented a personalized product search method, which falls into the second category. They modeled the long-term user preference through a latent space via jointly learning the representations of users, products and queries. This method also bears the aforementioned limitations. As far as we know, literature on the personalized product search is very sparse. To alleviate the problems faced by existing methods, we argue that integrating the relatively stable long-term and time-sensitive short-term user preferences can improve the search accuracy, and thus enhance the user's search experience.

However, an effective integration of the long- and short-term user preferences towards the personalized product search is non-trivial, due to the following facts: 1) Modeling the long- and short-term user preferences is complicated. The former contains multiple manifestations, e.g., \emph{consumption level}, \emph{preferred colors} and \emph{favorite brands}, and gradually evolves with the change of user's background, e.g., \emph{income}. Meanwhile, the short-term one is usually dynamic and time-sensitive, which can be easily affected by transient events. 2) For the user's current need expressed by a query, it is difficult to precisely identify which aspects in the long-term user preference are most relevant. For example, a \emph{T-shirt}'s design, instead of its \emph{price}, may impact more on the customers who can afford it. As to the short-term user preference, different products bought recently have different impacts on the user's next purchase. And 3) how to jointly encode the long- and short-term user preferences with the current query is another challenge.

To tackle these problems, we present an Attentive Long Short-Term Preference model (ALSTP for short), to perform the personalized product search by jointly integrating the current given query, the long- and short-term user preferences. Our model comprises three parts: 1) Attentive Short-term Preference Modeling. We establish the short-term user preference based on the products recently bought, rather than the full purchasing history. Subsequently, we leverage an attention mechanism to weight the constituent-wise impact of the short-term user preference given the current query. The short-term user preference is also used for updating the long-term user preference with a small rate. 2) Attentive Long-Term Preference Modeling. The long-term user preference is modeled by using a short-term session-based updating strategy based on the purchased products in chronological order. Similarly, an attention mechanism is used to weight the relevance of each factor in the long-term user preference with respect to the current query. And 3) we fuse the current query, the weighted short- and long-term user preferences to better represent the user's specific intention. Finally, a pairwise learning-to-rank method is used to get the best ranking list.

To verify the effectiveness of our proposed model, we conducted extensive experiments on four public Amazon product datasets. Experimental results demonstrate that our proposed ALSTP model yields better performance as compared to several state-of-the-art methods. It is worth mentioning that our model is applicable to many other scenarios, such as personalized movie search~\cite{park2007applying} and academic article search~\cite{tbahriti2006using}.

In summary, our main contributions of this paper are threefold:
\begin{itemize}[align=left,style=nextline,leftmargin=*,labelsep=\parindent,font=\normalfont]
  \item We present a neural network model for personalized product search by jointly integrating the long- and short-term user preferences, as well as the user's current query.
  \item We apply the attention mechanism to both the short- and long-term user preferences to weight the importance of different factors in both of them.
  \item We conducted extensive comparative experiments on four real-world Amazon datasets to thoroughly validate the effectiveness of our ALSTP approach. Moreover, we have released the codes and data to facilitate future  research in this direction\footnote{\href{https://github.com/guoyang9/ALSTP}{https://github.com/guoyang9/ALSTP.}}.
\end{itemize}

The rest of the paper is structured as follows. In Section~\ref{related_work}, we briefly review the related work. We detail the scheme of our model and its components in Section~\ref{model}. Experimental setup and result analysis are presented in Section~\ref{experiment}. We finally conclude our work and discuss the future directions in Section~\ref{conclusion}.

\section{Related Work}\label{related_work}
In this section, we focus on three categories of research directly related to our work: product search, personalized web search and deep learning techniques in information retrieval.
\subsection{Product Search}
Online product search has become an indispensable part of our daily life, as more and more users search and purchase products on the Internet~\cite{li2011towards}. There are basically two lines of research working on product search: click prediction and purchase prediction tasks. In view of clicks providing a strong signal of a user’s interest in an item, the methods in the first line~\cite{das2014commerce, parikh2011beyond, parikh2011user, goswami2011study, yu2014latent} focus on the click prediction of e-commerce users. For example, Yu et al.~\cite{yu2014latent} proposed a Latent Dirichlet Allocation (LDA) based method for diversified product search. Approaches in ~\cite{goswami2011study, chung2012impact} analyze the impact of images on user clicks. Recently, a challenge track named ``Personalized E-Commerce Search Challenge'' was sponsored by CIKM Cup 2016 to provide a unique opportunity for academia and industry researchers to test new ideas for personalized e-commerce search, which requires competitors to correctly predict the user clicks. Several groups~\cite{palotti2016learning, wu2017ensemble} reported their proposed ideas and experimental results.  However, the user clicks only indicate that the user may be interested in an item~\cite{chung2012impact}. Thus, relying on the click information to predict the user intention for the submitted query is dangerous when users accidentally click on wrong items or they are attracted by some unrelated items due to curiosity. In order to capture the user's main purpose for the submitted query, in this paper, we mainly focus on the purchase prediction.

Typically the e-commerce product inventory information is structured and stored in relational databases. As a result, associating the free-form user queries with these structured data (e.g., product entity specifications and detailed information)  can be hard for the search engine system. Some efforts have been dedicated to solving this problem~\cite{duan2013supporting, duan2013probabilistic, duan2015mining}. For instance, to fill the big gap between the keyword queries and the structured product entities,  Duan et al.~\cite{duan2013supporting, duan2013probabilistic} proposed a probabilistic retrieval model to optimize the ranking of product entities by mining and analyzing the product search log data. Although the language model is conceptually simple and computationally tractable, the major issue in the product search is that it overlooks the critical information from the entity space and the connection between the queries and product entities. Duan et al.~\cite{duan2015mining} noticed it and learned the query intent representation collectively in both the query space and the structured entity space. With the popularity of the online review platforms in recent years, researchers have attempted to extract information from reviews to represent products~\cite{cheng2018mmalfm, cheng2018aspect, chenwsdm18} by using representation learning techniques (e.g., word2vec~\cite{mikolov2013efficient, li2016topic, li2017enhancing}). Gysel et al.~\cite{van2016learning} introduced a latent semantic entity model to learn the distributed representations of words and entities (i.e., products) to solve the semantic mismatching problem in product search.

However, a good product search engine is far beyond the pure relevance between the queries and products. In the field of product search, besides the relevance, the user's personal preference for products is also critical to persuade users to purchase. Recently, Ai et al.~\cite{ai2017learning} proposed a personalized product search model and took into consideration the users' preference. They highlighted the fact that the purchasing behavior can be highly personal in online shopping and extended the model in~\cite{van2016learning} by mapping the user into the same latent space with the query and product. Concurrently, there is another work~\cite{guo2018multi} aiming to combine the visual preference and textual preference for product search. Nevertheless, they both ignored the short-term user preference and  assumed that the long-term user preference is stable.   In this paper, we emphasize the importance of integrating the long- and short-term user preferences with the current query in product search.
\subsection{Personalized Web Search}
Due to the information overload on the web and the diversity of user interests, personalization has been well-recognized as an important way to improve the search experience in web search~\cite{matthijs2011personalizing}. Instead of giving a complete review on the personalized web search, we mainly recap the related studies falling into the following two categories: short-term session-based and long-term profile-based web search, both being closely related to our work.

Short-term session-based approaches infer user's local preference by tracking user's behaviors in a short period, which is defined as a session. It involves two important problems, namely, \emph{how to determine the duration of a session} and \emph{how to learn the user preference in a session}. To solve these problems, Sriram et al.~\cite{sriram2004session} used temporal closeness and probabilistic similarities between queries to determine the duration of a session; Daoud et al.~\cite{daoud2009session, daoud2008learning} leveraged a predefined ontology of semantic concepts to construct the short-term user profile; methods in~\cite{white2010predicting, ustinovskiy2013personalization} determine the extent of using context (e.g., prior queries and clicks) and investigate the combination of the query and the context to model the user intention; and Shen et al.~\cite{shen2005implicit} captured users' interests by analyzing the immediate contexts and implicit feedback.

Long-term profile-based approaches model the long-term user preference based on the user's overall search logs. Approaches in ~\cite{tan2006mining, sontag2012probabilistic} apply a probabilistic statistical language modeling technique to discover the relevant context of the current query from the search history. Richardson et al.~\cite{richardson2008learning} analyzed the long-term query logs to learn more about the user behavior. Matthijs et al.~\cite{matthijs2011personalizing} adopted NLP techniques, such as web page structure parsing, term extraction and part of speech tagging, to extract noun phrases to mine the user profile. Bennett et al.~\cite{bennett2012modeling} leveraged a simple time decay mechanism to combine the long- and short-term user profiles and pointed out that the long-term behavior provides substantial benefits at the start of a search session while the short-term session behavior contributes the majority of gains in an extended search session.

The decision mechanism that underlies the process of buying a product is different from locating relevant documents or objects~\cite{li2011towards}. In product search, simply returning something relevant to the user's submitted query may not lead to the purchasing behavior. For example, a returned relevant product which is far beyond a user's consumption capability would not be purchased by this user. Fortunately, such attributes (e.g., consumption capability) of users can be reflected in the user's long-term profile. Meanwhile, short-term behaviors usually provide evidence of user's local purchasing inclinations. Therefore, it is necessary to consider both the long- and short-term user preferences in the product search. Approaches in personalized web search scarcely consider such problems, and thus cannot be directly applied to the product search.
\subsection{Deep Learning in Information Retrieval}
With the success of deep learning techniques in many domains, such as Computer Vision, Natural Language Processing and Speech Recognition, researchers also applied them to Information Retrieval (IR) and Recommender Systems, and achieved promising performance.

\textbf{Deep Learning in IR.} The deep learning techniques in IR can be roughly classified into two categories: the semantic-oriented and the relevance-oriented. The former mainly concerns the semantic matching between the queries and products. The basic idea under this category is that they first process the text of both the queries and documents, and then build the interaction between them. For example, Huang et al.~\cite{huang2013learning} leveraged a deep neural network (DNN) to project the queries and documents into a common low-dimensional space, and then calculate the relevance of each document with respect to the given query in this space. Shen et al.~\cite{shen2014learning} introduced a convolutional neural network (CNN) to capture the fine-grained contextual structures of entities (queries and documents). Other semantic matching methods like ARC-1~\cite{hu2014convolutional}, ARC-2~\cite{hu2014convolutional}, Match-SRNN~\cite{wan2016match}, MatchPyramid~\cite{pang2016study} and~\cite{severyn2015learning} also share the similar idea. Guo et al.~\cite{guo2016deep} pointed out that the aforementioned methods only focus on the semantic matching between the queries and documents. However, ad-hoc retrieval task is mainly about the relevance matching. They proposed a jointly deep architecture to model three key factors of relevance matching. Nevertheless, the proposed model failed to explicitly model the relevance generation process and capture the important IR characteristics. Pang et al.~\cite{pang2017deeprank} extended the method in~\cite{guo2016deep} to better capture the intrinsic relevance and simulate the human judgment process. In addition, Borisov et al.~\cite{borisov2016neural} introduced a neural click model to better understand the user browsing behaviour for the web search.

\textbf{Deep Learning in Recommendation.} However, the aforementioned approaches have not taken personalization into consideration, yet it is vital for the product search. In fact, deep learning has been successfully applied in recommender systems to model the user preference~\cite{li2015deep, he2017neural, cheng20183ncf}. In particular, the attention mechanism is usually adopted in these systems to model the user's preference more accurately. Attention mechanisms have shown its efficiency in various tasks such as image captioning~\cite{xu2015show, chen2017sca}, visual question answering~\cite{xu2016ask, yang2016stacked}, machine translation~\cite{bahdanau2014neural} and information retrieval~\cite{li2017neural, chen2017attentive, xiao2017attentional, phan2017neupl}. A survey on the attention mechanism on those domain is out of the scope of this paper. Here we only focus on the attention mechanism on user preference modeling. To name just a few, in the field of recommender systems, Li et al.~\cite{li2017neural} introduced a neural attentive model to the session-based recommendation. They explored a hybrid encoder with an attention mechanism to model user's sequential behaviors and capture their emphases in the current session. Chen et al.~\cite{chen2017attentive} proposed a two-level attention mechanism (i.e., item- and component-level) for the multimedia recommendation. Based on this, a user-based collaborative filtering approach to modeling user preference on both the item-level and component-level is seamlessly applied on implicit feedback. Xiao et al.~\cite{xiao2017attentional} improved Factorization Machine (FM) by discriminating
the importance of different feature interactions.

In this paper, we apply deep learning techniques to model the user preferences in the product search. In particular,  we use the attention mechanism to model the relation among the query, long- and short-term user preferences to better represent the user's search intention (as we take user's relatively stable purchasing bias and recent purchasing intention into consideration). To the best of our knowledge, we are the first to apply the attention mechanism in both the long- and short-term user preferences to precisely capture the user's search intention. The experimental results demonstrate that the attention mechanism is essential to improve the retrieval performance.

\section{Our Proposed Model}\label{model}
\subsection{Preliminaries}
\textbf{Research Problem.}
Given a query $q$ of a user $u$, our goal is to return a ranked list of products based on their matching scores with respect to $u$'s current search intention. It is difficult for a query $q$, typically consisting of a few keywords, to precisely describe the user's search intention. Towards this end, we augment the current query by integrating the user's \emph{long-term preference} and \emph{short-term preference}. Consequently, the key problem is divided into two subproblems: 1) how to model the long- and short-term user preferences; and 2) how to effectively integrate them with the current query to capture the user's search intention.

\textbf{Model Intuition.} Before describing the proposed model, we would like to introduce the underlying intuitions.
We start with the definition of two key concepts.
User's \textbf{long-term preference (LTP)} refers to user's general interests or inclinations on products over a long time period, such as user's favored \emph{clothing style, sports} and \emph{genres of movies}. It is inherently correlated with the user's background (e.g., \emph{gender, age, career,} or \emph{consumption capability}).  Besides, the long-term user preference would change only gradually at a rather slow rate. On the contrary, the
\textbf{short-term preference (STP)} expresses user's local shopping intention, which is usually situation- or event-dependent, such as \emph{weather} (e.g., \emph{winter} or \emph{summer}) and \emph{recent events} (e.g., \emph{wedding}).  For example, users often purchase \emph{T-shirt} and \emph{shorts} together, and buy accessories (e.g., \emph{mouse} or \emph{keyboard}) after purchasing a device (e.g.,
\emph{computer}).  Usually, the recent purchasing behaviors provide some evidence of the current short-term user preference.  In summary, LTP reflects user's overall interests and is slowly evolving (due to the changing backgrounds); and in contrast, STP represents user's local purchasing inclinations. Therefore, we model LTP using user's overall transaction history with a slow updating strategy and model STP based on user's previous purchasing behaviors in a recent short time window.

Let us use an example to illustrate why the integration of LTP and STP can help us to infer the user's purchasing intention. Given a  \emph{``phone case''} as the query of a user, which is quite ambiguous, since \emph{cases} could be of different \emph{colors}, \emph{styles}, and for different \emph{phones}. From LTP, we could know the user's preferred \emph{price}, \emph{color} and \emph{style}; while from STP we might know the specific model of the case, e.g., \emph{the user recently bought an iPhone 7}. Meanwhile, we should notice that LTP covers the user preference for products from different aspects, which is not only limited to the information related to the current query, such as user preferences over other products. Similarly, STP could also contain information quite different from the current search intention, i.e., not all products purchased recently are directly related to the current query. Therefore, identifying the  specific parts in LTP and STP with respect to the current query could capture the user's search intention more accurately and avoid the risk of ``query drift''~\cite{zighelnic2008query}. To achieve the goal, before fusing LTP, STP and the current query, we design two attention mechanisms for LTP and STP, respectively.

Formally, let $\mathcal{P}_u = \{p_1, p_2, ...,p_N\}$ denote all of the user $u$'s purchased products list in a chronological order, and $\mathcal{Q}_u = \{q_1, q_2, ..., q_N\}$ denote the corresponding query list.
As queries and the purchased products correspondingly appear in pairs (i.e., <$q_i$, $p_i$>)\footnote{Notice that in real scenarios, a query $q_i$ could lead to the purchase of several products (e.g., $p_l$ and $p_{m}$). In our model , it can be regarded as two search processes as <$q_i$, $p_l$> and <$q_i$, $p_m$>.} and both can indicate user's local search intention or preference, we will use them exchangeably to describe the user preferences.  In our model, we infer user $u$'s STP for a query $q_n$  by referring to his/her recent $m$ queries (i.e., $\{q_{n-m}, ..., q_{n-1}\}$) and corresponding products (i.e., $\{p_{n-m}, ..., p_{n-1}\}$) (detailed in Sect.~\ref{sect:ASTPM}). For the LTP modeling, the first several purchased products are used to initialize LTP, which is then gradually updated based on STP with a small rate (detailed in Sect.~\ref{sect:ALTPM}). The main notations used in this paper are summarized in Table~\ref{notation}.
\begin{table}
\caption{Main notations used in this paper.}\label{notation}
\begin{tabular}{c|l}
\hline
$\mathcal{Q}_u$ & user $u$'s issued query list in chronological order\\
\hline
$\mathcal{P}_u$ & user $u$'s purchased product list corresponding to his/her query list \\
\hline
$\bm{q}$, $\bm{p}$ &  latent representation of query and product, respectively \\
\hline
$\hat{\bm{q}}$, $\hat{\bm{p}}$ & transformed latent representation of query and product, respectively \\
\hline
$\bm{g}$ & latent representation of the long-term user preference  \\
\hline
$\bm{h}_i$, $\bm{h}_{n-1}$ & the hidden representation of the time step $i$ and the final representation from $GRU$ \\
\hline
$\bm{c}^l$, $\bm{c}^g$ & the attentive short- and long-term user preference representations, respectively \\
\hline
$\bm{c}$, $\bm{c}_l$& the unified user intent representation and its $l$-th layer projection, respectively \\
\hline
$\bm{W}$, $\bm{b}$ & weight matrix and bias vector, respectively \\
\hline
$\bm{v}$ & the weight vector for computing the short-term attention score \\
\hline
$\phi$ & activation function \\
\hline
$k$ & the dimension of both queries and products \\
\hline
$m$ & the short-term window size \\
\hline
$\beta$ & the long-term user preference updating rate \\
\hline
$a^{l}_{i}$, $\alpha^{l}_{i}$ & the short-term attention score before and after normalization, respectively \\
\hline
$a^{g}_{i}$, $\alpha^{g}_{i}$ & the long-term attention score before and after normalization, respectively \\
\hline
<$q$, $p$> & query product pair \\
\hline
\end{tabular}
\end{table}

\textbf{Query and Product Representation.} The PV-DM model~\cite{le2014distributed} is adopted for the latent vector representation learning of queries and products. PV-DM is an unsupervised method to learn the continuous distributed vector representations for textual documents. It takes word sequence into consideration and can preserve the semantic features of words. PV-DM takes text documents as inputs and outputs their vector representations in a latent semantic space. In our context, the reviews of products and textual queries are mapped into the same latent space via PV-DM to learn their vector representations~\footnote{The inputs to PV-DM model are each products' reviews and textual queries (In order to keep the sematic relationship between queries and reviews). After the unsupervised training of PV-DM, each of these inputs has its own output, e.g., equal-length vector, we take this vector as the representation of each product and query. Afterwards, when a new query comes, we can easily obtain its vector representation by inputting this textual query to the trained PV-DM model. And then a ranked product list can be returned to this query.}, which are then used as inputs in our model. Without loss of generality, we assume the vector representations of the long-term user preference, products and queries have the same dimension $k$. Let $\bm{p}_i \in \mathbb{R}^k$ be the  product $p_i$'s representation and $\bm{q}_{j} \in \mathbb{R}^k$ be the query $q_{j}$'s representation.
\begin{figure*}
\includegraphics[width=0.9\textwidth]{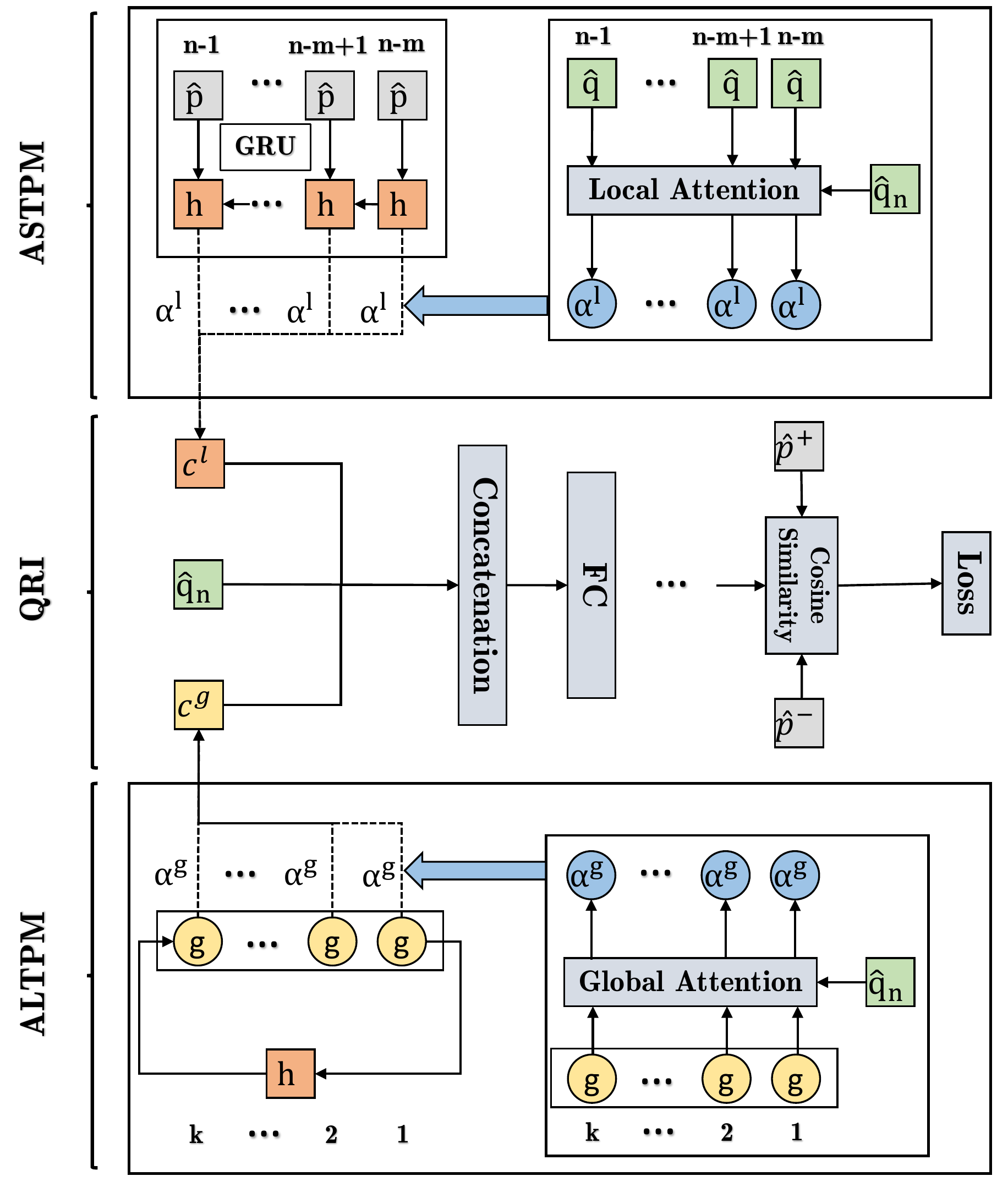}
\caption{The architecture of our proposed ALSTP framework. Symbol $\hat{\bm{p}}$ denotes a product's vector representation after the transformation, and  $\hat{\bm{q}}$ is a query's vector representation. The top part is the Attentive Short-Term Preference Modeling (ASTPM), which is useful to obtain the attentive short-term user preference $\bm{c}^l$. The bottom part Attentive Long-Term Preference Modeling (ALTPM) is to learn the attentive long-term user preference $\bm{c}^g$. Finally, the middle part Query Representation Integration (QRI) fuses these two with the current query to better represent user's shopping intention and leverages a comparative learning method to obtain the final loss.}
\label{fig:framework}
\end{figure*}
\subsection{Model Description}
\subsubsection{Overall Framework} Figure~\ref{fig:framework} shows the structure of the proposed ALSTP model. The model comprises three parts:
 \emph{Attentive Short-Term Preference Modeling (ASTPM)}, \emph{Attentive Long-Term Preference Modeling (ALTPM)},  and \emph{Query Representation Integration (QRI)}.
\begin{itemize} [align=left,style=nextline,leftmargin=*,labelsep=\parindent,font=\normalfont]
 	\item \textbf{ASTPM.}  This part is to extract the short-term user preference based on $m$ immediately preceding queries and their corresponding purchased products, and learn the relations between the current query and those queries by an attention mechanism.
 	\item \textbf{ALTPM.} This part models the long-term user preference via a gradually updating strategy based on the historical purchased products in chronological order, and it uses an attention mechanism to associate the current query with the long-term user preference.
 	\item \textbf{QRI.} With the attentive modeling of the short- and long-term user preferences, the last part integrates them with the current query to represent the user's current search intention.
\end{itemize}
In the following, we detail the three parts sequentially.

\subsubsection{Attentive Short-Term Preference Modeling} \label{sect:ASTPM}
For each query $q$ and product $p$, we map them into the same latent space through a fully connected layer as,
\begin{equation}\label{eq:trans1}
\hat{\bm{p}} = \phi(\bm{W}_c\bm{p} + \bm{b}_c),
\end{equation}
\begin{equation}\label{eq:trans2}
\hat{\bm{q}} = \phi(\bm{W}_c\bm{q} + \bm{b}_c),
\end{equation}
where $\bm{q}$ and $\bm{p}$ are vector representations of the query $q$ and product $p$ obtained from PV-DM (as described in the last subsection).  $\bm{W}_{c}$ $\in$ $\mathbb{R}$$^{k{\times}k}$ and $\bm{b}_c$ $\in$ $\mathbb{R}$$^{k}$ are the weight matrix and bias vector, respectively. $\hat{\bm{p}}$ and $\hat{\bm{q}}$ are the product and query representations after transformation, respectively.  $\phi$(\textbullet) is the activation function. In our implementation, \emph{ELU}~\cite{clevert2015fast} is adopted\footnote{The activation functions could be \emph{sigmoid}, hyperbolic tangent (\emph{tanh}), rectified linear unit (\emph{ReLU}) , leaky \emph{ReLU} or exponential linear unit (\emph{ELU}).  We tested all those activation function in experiments. And the \emph{ELU} function achieved better performance.}.

Given the current query $q_n$, considering that the recently purchased products may relate to the current search intention, several recent queries and the products purchased accordingly are used to capture the short-term user preference. Specifically, let $m$ be the time window size\footnote{As users' purchasing transactions are usually too sparse to learn, for example, some users have no purchase transactions in a specific time period (e.g., a month). Thus, we choose to use a short-term time window (e.g., the recently purchased four products) to learn the short-term user preference.}.  To model the influence imposed by the previous queries (or products) on the next purchase, we adopt a Recurrent Neural Network  (\emph{RNN}) model equipped with Gated Recurrent Units (\emph{GRU})~\cite{cho2014properties} to model the short-term user preference, as it has been successfully applied in the session-based product recommendation~\cite{hidasi2015session, li2017neural}.

\textbf{GRU} was proposed in~\cite{cho2014properties}. The activation of GRU is a linear interpolation between the previous activation $\bm{h}$$_{t-1}$ and the candidate activation $\hat{\bm{h}}$$_t$,
\begin{equation}\label{}
\bm{h}_t = (1 - z_t)\bm{h}_{t-1} + z_t\hat{\bm{h}}_t,
\end{equation}
where the update gate z$_t$ is given by,
\begin{equation}\label{}
z_t = \sigma(\bm{W}_z\bm{x}_t + \bm{U}_z\bm{h}_{t-1}).
\end{equation}
The candidate activation function $\hat{\bm{h}_t}$ is computed as
\begin{equation}\label{}
\hat{\bm{h}}_t = \emph{tanh}[\bm{W}\bm{x}_t + \bm{U}(\bm{r}_t \odot \bm{h}_{t-1})],
\end{equation}
where the rest gate $\bm{r}$$_t$ is given by
\begin{equation}\label{}
\bm{r}_t = \sigma(\bm{W}_r\bm{x}_t + \bm{U}_r\bm{h}_{t-1}).
\end{equation}
The \emph{RNN} module takes the recent $m$ product representations [$\hat{\bm{p}}$$_{n-m}$, $\cdots$, $\hat{\bm{p}}$$_{n-1}$] as inputs. We thus can obtain a set of high dimensional hidden representations [$\bm{h}$$_{n-m}$, $\cdots$, $\bm{h}$$_{n-1}$], denoting the short-term user preference. The hidden state of RNN is initialized by the long-term user preference $\bm{g}$, which is introduced in the next subsection.

\begin{figure*}
\includegraphics[width=0.8\textwidth]{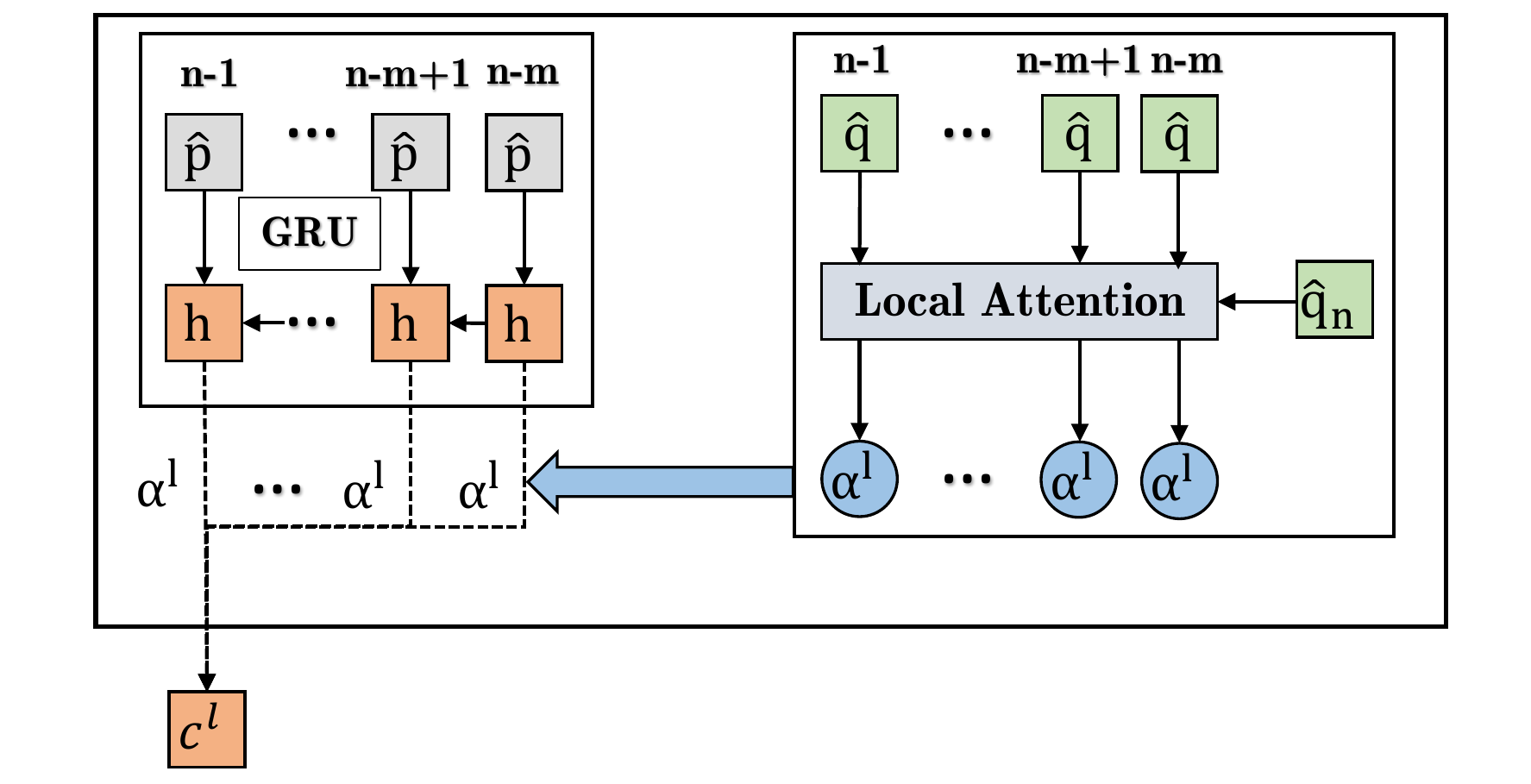}
\caption{Illustration of Attentive Short-Term Preference Modeling.}
\label{fig:short-term}
\end{figure*}

\textbf{Attention Mechanism.} Notice that the previous queries (or purchased products) are not all closely relevant to the current query (targeted products). For example, previous queries about \emph{mouse} and \emph{keyboard} are more relevant than the \emph{phone} with the current query of \emph{display screen}. Considering that these previous queries have different relevance levels to the current query, we propose to use an attention mechanism to capture their relevance scores or attentions based on the representations of queries. Let $\hat{\bm{q}}$$_{n}$ be the representation of the current query and [$\hat{\bm{q}}$$_{n-m}$, $\cdots$, $\hat{\bm{q}}$$_{n-1}$]  denote the representations of the previous query list corresponding to the previous product list of [$\hat{\bm{p}}$$_{n-m}$, $\cdots$, $\hat{\bm{p}}$$_{n-1}$]. As shown in Fig~\ref{fig:short-term}, we apply a two-layer neural network to estimate the attentions [$a{^l_{n-m}}$, $\cdots$, $a{^l_{n-1}}$] based on $\hat{\bm{q}}$$_{n}$ and [$\hat{\bm{q}}$$_{n-m}$, $\cdots$, $\hat{\bm{q}}$$_{n-1}$] ,
\begin{equation}\label{eq:short}
a{^l_j} = \bm{v}^{\bm{T}}\phi(\bm{W}{^l_0}\bm{q}_n + \bm{W}{^l_1}\bm{q}_j + \bm{b}^l),
\end{equation}
where $\bm{W}{^l_0}$ $\in$  $\mathbb{R}$$^{f{\times}k}$ and $\bm{W}{^l_1}$ $\in$ $\mathbb{R}$$^{f{\times}k}$ ($f$ is a scalar hyper-parameter) are matrix parameters of the first layer, $\bm{b}^l$ $\in$ $\mathbb{R}^k$ is the bias, $\bm{v}$ $\in$ $\mathbb{R}$$^{f{\times}1}$ is the vector parameter of the second layer, and $\phi$ is the activation function $ELU$.
Subsequently, the attention weights are further normalized by applying a \emph{softmax} function:
\begin{equation}\label{eq8}
\alpha{^l_j} = \frac{\emph{exp}(a{^l_j})}{\sum_{i=n-m}^{n-1}{\emph{exp}(a{^l_i})}}.
\end{equation}

Accordingly, the previously purchased products of these queries contribute differently to the final short-term user preference modeling. The role of $\alpha{^l}$ is to determine which products bought before should be emphasized or ignored to capture the attentive short-term user preference. In this way, the attentive short-term user preference is modeled as a linear combination of the short-term user preference weighted by the corresponding attentions obtained by Eqn.($\ref{eq8}$),
\begin{equation}\label{eq:attshort}
\bm{c}^l = \sum_{i=n-m}^{n-1}{\alpha{^l_i}\bm{h}_i},
\end{equation}
where $\bm{c}$$^l$ is the final representation of the attentive short-term user preference. It is expected that $\bm{c}$$^l$  captures the information in the short-term user preference related to the current query. The \emph{RNN}-based module can update the short-term user preference immediately and effectively.

\subsubsection{Attentive Long-Term Preference Modeling} \label{sect:ALTPM}
By contrast, the long-term user preference is relatively stable and updates slowly. Although it is modeled based on the overall purchasing history of the user, in our model, we initialize it by a set of products purchased at early times and then update it by the products bought subsequently.  Let $\bm{g} \in \mathbb{R}^k$ denote the long-term user preference, which corresponds to latent factors representing user's inherent purchasing bias (e.g., favorite colors, preferred fashion design in \emph{Clothing} dataset). For simplicity, we use the first $m$ products to set it up, and then update it with every $m$ products purchased (evolving gradually). In this setting, $\bm{g}$ is updated  based on a session of $m$ products, which is also the window size of modeling the short-term user preference. Therefore, $\bm{g}$ is updated via:
\begin{equation}\label{eq13}
\bm{g} = (1 - \beta)\bm{g} + \beta\bm{h}{^\prime_{n-1}},
\end{equation}
where $\bm{h}$${^\prime_{n-1}}$ is the final hidden representation of the short-term user preference based on the previous $m$ products, and $\beta$ is a hyper-parameter denoting the updating rate.

\begin{figure*}
\includegraphics[width=0.8\textwidth]{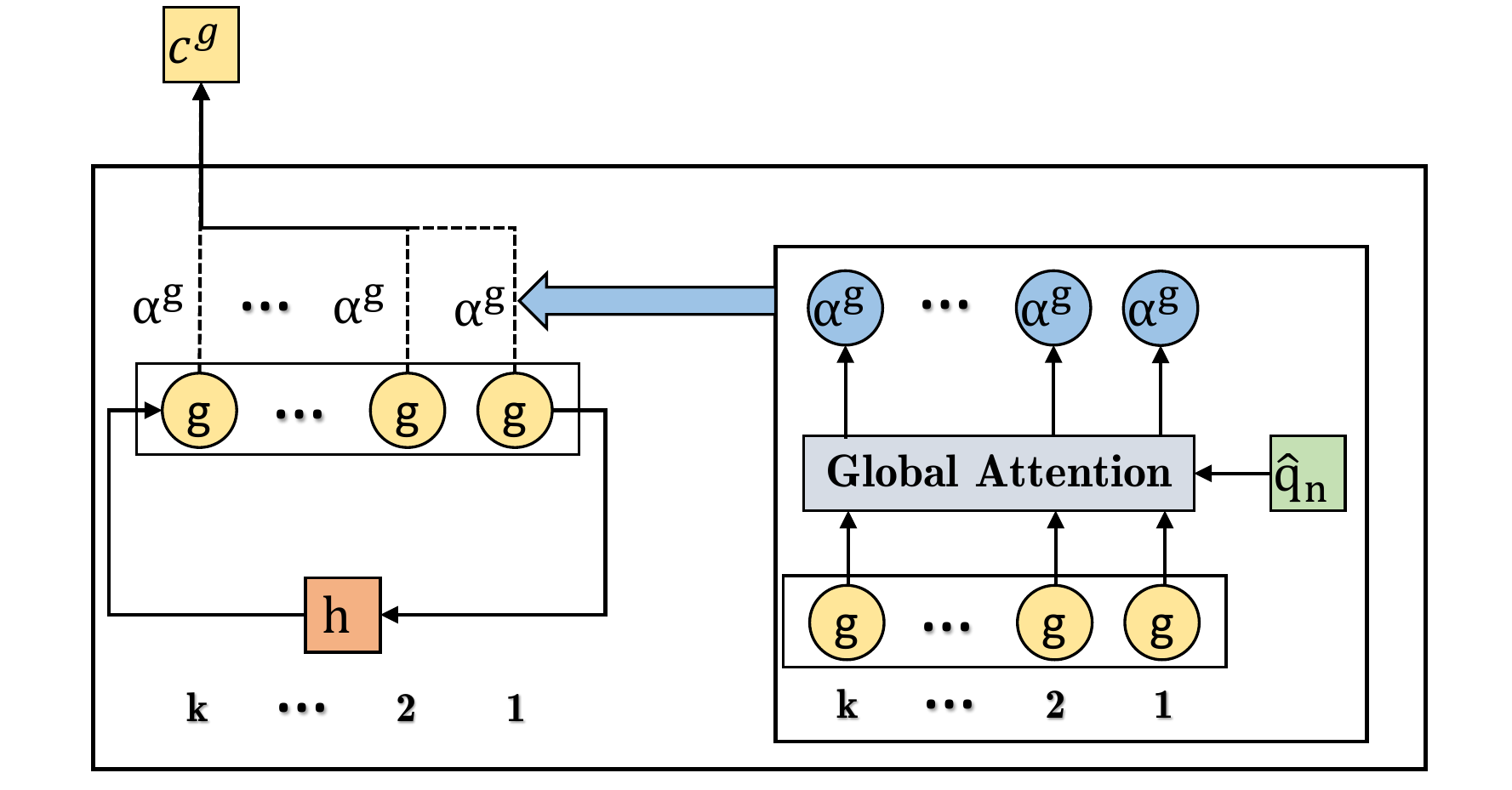}
\caption{Illustration of Attentive Long-Term Preference Modeling.}
\label{fig:long-term}
\end{figure*}

\textbf{Attention Mechanism.} We argue that users can perform distinctively towards different queries based on their specific long-term preference. Therefore, the attention mechanism for long-term user preference is to find out which factors in $\bm{g}$ are more important for the current query. For example, if a user can afford most of the laptops on the e-commerce website, and then a machine's performance or appearance, instead of its price, will affect the user's final purchase decision more. Because of that, a good product search engine should learn how to emphasize the influential factors in the long-term user preference and depress the inessential ones. Thus, in this paper, similar to the attentive short-term preference modeling, we also use an attention mechanism to measure the relevance between different aspects in the long-term user preference and the current query,
\begin{equation}\label{eq10}
a{^g_j} = g_j\bm{w}{^g}\hat{\bm{q}}_n + b^g,
\end{equation}
where $g_j$ is the $\mathit{j}$-th element of the long-term user preference, $\bm{w}{^g}$ $\in$ $\mathbb{R}$$^{1{\times}k}$ is the weight parameter, and b$^g$ is the bias. Fig~\ref{fig:long-term} illustrates the main idea of the computation of these attention weights. The final attention $\bm{\alpha}$$^g$ = [$\alpha$${^g_1}$, $\cdots$, $\alpha$${^g_k}$] is obtained by normalization via a \emph{softmax} function,
\begin{equation}\label{eq11}
\alpha{^g_j} = \frac{\emph{exp}(a{^g_j})}{\sum_{i=1}^{k}{\emph{exp}(a{^g_i})}}.
\end{equation}

Finally, we compute the element-wise product of the long-term user preference and the attention weights by,
\begin{equation}\label{eq12}
\bm{c}^g = \bm{g} \odot \bm{\alpha}^g,
\end{equation}
where $\bm{c}$$^g$ is the representation of the attentive long-term  user preference. It is expected to capture the information in the long-term user preference related to the current query.

\subsubsection{Query Representation Integration}
\begin{figure*}
\includegraphics[width=0.7\textwidth]{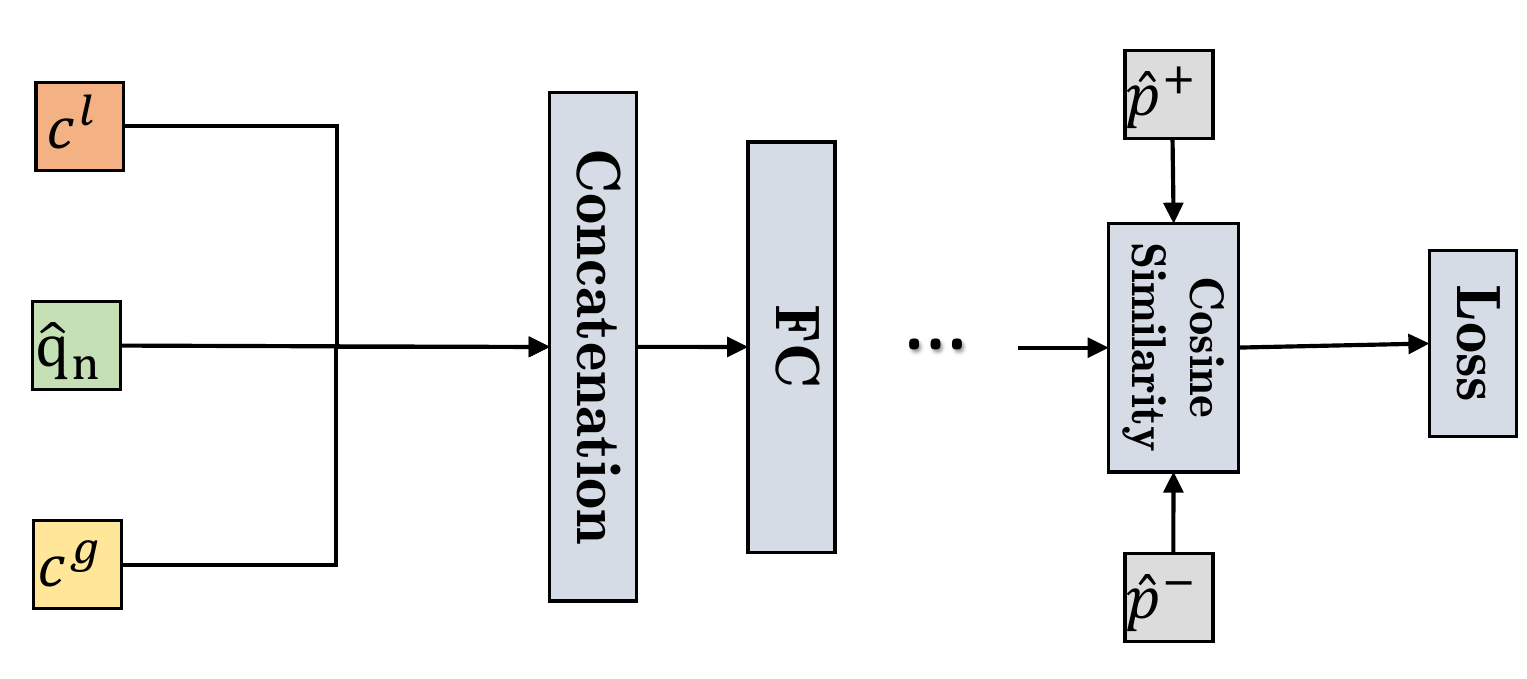}
\caption{Illustration of Query Representation Integration.}
\label{fig:merge}
\end{figure*}

With respect to the query $q$, $\bm{c}_g$ has summarized the current attentions on the long-term user preference; and $\bm{c}_l$  adaptively captures the relevant purchasing intentions in the short-term user preference. Besides, the query is the most direct representation of the user's current need. The basic idea of this work is to learn a product search model that considers both the short- and long-term user preferences to identify the current purchasing intention. Based on this idea, as shown in Fig~\ref{fig:merge}, we fuse the above three parts into a unified representation $\bm{c}$ by concatenation,
\begin{equation}\label{eq14}
\bm{c} = [\bm{q}_n; \bm{c}^l; \bm{c}^g].
\end{equation}

We then map this unified representation into the same latent space with the products for direct matching. To model the interactions between these three parts and obtain better fusion features, we refer to Deep Neural Networks (DNNs), which introduces multi-layer of non-linear interactions and has been proven to be very effective in the feature fusion tasks ~\cite{zhang2014start}. Specifically, the fully connected layers are:
\begin{equation}\label{eq:split}
\begin{split}
  \bm{c}_1 &= \phi (\bm{W}_1\bm{c} + \bm{b}_1),\\
  \bm{c}_2 &= \phi (\bm{W}_2\bm{c}_1 + \bm{b}_2), \\
  ......,\\
  \bm{c}_L &= \phi (\bm{W}_L\bm{c}_{L-1} + \bm{b}_L),
\end{split}
\end{equation}
where $\bm{W}_l$ and $\bm{b}_l$ denote the weight matrix and bias vector for the $l$-th fully connected layer, respectively. $\phi(\cdot)$ is the activation function $ELU$. Besides, our network structure follows a tower pattern, where the bottom layer is the widest and each
successive layer has a smaller number of neurons~\cite{he2017neural}. Ultimately, the output from the last layer $\bm{c}_L$ has the
dimension of $k$, equal to the query laten representation size.

With the current upgraded query representation $\bm{c}_L $ and product representation $\hat{\bm{p}}$, the relevance score between the query and product $p_j$ is computed as,
\begin{equation}\label{eq16}
s_j = \varphi(\bm{c_L}, \hat{\bm{p}}_j),
\end{equation}
where $\varphi$ is a distance function that can be cosine similarity, dot product, Euclidean or Manhattan distances. In our experiments, we observed that cosine similarity yields relatively better performance. After that, all products are ranked in a descending order based on their relevance scores with respect to the current query. The top ranked products are returned to users.

\subsection{Optimization}
We use a pair-wise learning method to train the model. All the observed query-product pairs are treated as positive pairs. For each positive pair $(q, p^+)$, we randomly sample $N_s$ negative products $p^-$. A query with a positive product and a negative product constructs a triplet  $(q, p^+, p^-)$ for training. The BPR loss is adopted, which is a pairwise method and has been widely used in the recommender systems towards personalized ranking~\cite{rendle2009bpr}.
\begin{equation}\label{eq17}
Loss(\bm{c}_L; \hat{\bm{p}}^+; \hat{\bm{p}}^-) = -\emph{log}(\sigma(s_{p}^+ - s{_{p}^{-}})) + \lambda(\|\Theta\|^2),
\end{equation}
where $\hat{\bm{p}}$$^+$ and $\hat{\bm{p}}$$^-$ denote the representations of the positive and negative samples, respectively. And s$_{p}^+$ and s${_{p}^{-}}$ are the corresponding positive and negative sample scores. $\lambda$ is the $\ell_2$ regularization hyper-parameter, $\sigma$ is the sigmoid function, and $\Theta$ denotes all the parameters in our model. To optimize the objective function, we employ the stochastic gradient descent (\emph{SGD}), a universal solver for optimizing the neural network models. The steps for training the network are summarized in Algorithm~\ref{alg}.

\begin{table}
\centering{
\begin{minipage}[c]{0.8\textwidth}
\begin{algorithm}[H]
\caption{Attentive Long- and Short-term Preference Modeling.}
\label{alg}
\SetKwInOut{Input}{Input}
\SetKwInOut{Output}{Output}
\Input{(Query, Product) latent representation pairs in chronological order.}
\Output{Latent representation of the long-term user preference and all the learned parameters.}
Initialize all parameters $\Theta$ with xavier~\cite{glorot2010understanding} distribution;

\For{each user u}{
Initialize long-term user preference $\bm{g}$ with zeros;

\For{each $m$ (query, product) pairs}{
Transform them into the same latent space according to Eqn.~\ref{eq:trans1} and Eqn.~\ref{eq:trans2};

Compute the product latent representations [$\hat{\bm{p}}$$_{n-m}$, $\cdots$, $\hat{\bm{p}}$$_{n-1}$];

Compute  local attention weights $\bm{\alpha^{l}}$ according to Eqn.~\ref{eq:short} and Eqn.~\ref{eq8};

Compute attentive short-term preference $\bm{c}^l$ according to Eqn.~\ref{eq:attshort};

Compute global attention weights $\bm{\alpha^{g}}$ according to Eqn.~\ref{eq10} and Eqn.~\ref{eq11};

Compute attentive long-term preference $\bm{c}^g$ according to Eqn.~\ref{eq12};

Compute final query representation $\bm{c}_L$ according to Eqn.~\ref{eq:split};

\For{each sampled negative product}{
Compute the final loss according to Eqn.~\ref{eq17};

update $\Theta$;
}
\If{$m$ products purchased}{
Update $\bm{g}$ according to Eqn.~\ref{eq13};
}
}
}
\end{algorithm}
\end{minipage}
}
\end{table} 

\section{Experiments}\label{experiment}
We conducted extensive experiments on several datasets to thoroughly justify the effectiveness of our model. In particular, our experiments mainly answer the following research questions:

\begin{itemize} [align=left,style=nextline,leftmargin=*,labelsep=\parindent,font=\normalfont]
	\item \textbf{RQ1:} Can our model outperform the state-of-the-art product search methods? (Sect.~\ref{sect:comp})
	\item \textbf{RQ2:} How does each component in our model affect the final performance, including the short-term user preference and long-term one? (Sect.~\ref{sect:ablation})
    \item \textbf{RQ3:} Are the attention mechanisms and long-term user preference updating strategy helpful to the final product search results? (Sect.~\ref{sect:ablation})
	\item \textbf{RQ4:} How do the important parameters influence the performance of our model, including the window size of the short-term user preference modeling and the embedding size of the user preferences and product representations? (Sect.~\ref{sect:parameters})
	\item \textbf{RQ5:} Can our attentive short-term preference module capture  the impact of the previous transactions on the current purchasing intention? (Sect.~\ref{sect:examples})
\end{itemize}
In the following, we first described the experimental setup. Then the performance comparisons of the proposed model ALSTP with several state-of-the-art product search methods are reported. Thereafter, we analyzed the utility of each model component by the performance comparisons among different variants of our ALSTP model, followed by the analysis of the effectiveness of the attention mechanism and the long-term user preference updating strategy. We then studied the influence of the two important parameters involved in our model and finally illustrated the effectiveness of the attention mechanism with some case studies.

\subsection{Experimental Setup}

\subsubsection{Dataset}
We experimented on the public Amazon product dataset\footnote{\href{http://jmcauley.ucsd.edu/data/amazon/}{http://jmcauley.ucsd.edu/data/amazon/.}}. The dataset contains product reviews and metadata from Amazon including hundreds of millions of reviews spanning from May, 1996 to July, 2014. It is organized into 24 product categories. In our experiments, we adopted the 5-core version provided by McAuley et al.~\cite{mcauley2015inferring}, whereby the remaining users and products have at least 5 reviews, respectively. Besides, we selected four categories with different sizes: \emph{Phones}, \emph{Toys},  \emph{Clothing}, and \emph{Electronics}. Following the strategy in~\cite{mcauley2015image, he2016ups}, we extracted the users' product purchasing behaviors based on their reviews, i.e., the products they reviewed are the ones they purchased. Our model uses the previously purchased products in a neighboring window size to model the short-term user preference. In order to study the influence of the window size on our model, we further filtered the dataset to make sure each user has at least 10 purchased products (i.e., 10 reviews).  And the words with low frequency were removed in our experiment. The basic statistics of these processed subsets are shown in Table~\ref{tab:statistic}. Note that the number of feedback equals to that of the query-product pairs.

\begin{table}
	\caption{Statistics over the evaluation datasets.}
	\label{tab:statistic}
	\begin{tabular}{r|c|c|c|c|c|c}
		\hline
		\multirow{2}{*}{Datasets}&\multirow{2}{*}{\#Users}&\multirow{2}{*}{\#Products}&
        \multirow{2}{*}{\#Feedback}&\multirow{2}{*}{\#Queries}&Avg. Query&Avg. Review\\
        &&&&&                                               Length    &Length     \\
		\hline
		Phones        & 3,216  & 9,018  & 47,139  & 107   & 6.61 & 135.39\\
		Clothing      & 5,200  & 20,424 & 72,142  & 1,325 & 7.64 & 35.88\\
		Toys          & 4,188  & 11,526 & 74,423  & 371   & 7.05 & 103.09\\
		Electronics   & 45,225 & 61,918 & 773,502 & 839   & 6.89 & 160.83 \\
		\hline
	\end{tabular}
\end{table}

\textbf{Query Extraction.}
As Rowley described in ~\cite{rowley2000product}, a typical scenario of user searching a product is to use \emph{a producer's name, a brand or a set of terms which describe the category of the product} as the query in retrieval. Based on this observation and following the strategy of~\cite{van2016learning, ai2017learning}, for each product a user purchased,  we extracted the corresponding search query from the categories to which the product belongs. The extraction of the textual representation for queries of the topics based on the categories is detailed as follows. We first extracted the category information for each product from its metadata. And then we concatenated the terms from a single hierarchy of categories to form a string of topics. Finally, we removed the punctuation, stop words and duplicate words from this topic string. To eliminate the duplicate words, we maintained the terms from sub-categories in view of these terms carrying more specific information compared to their parent-categories. For example, in the dataset of \emph{Phones}, the extracted query for \emph{Cell Phones \& Accessories $\rightarrow$ Accessories $\rightarrow$ Batteries $\rightarrow$ Internal Batteries} would be ``\emph{cell phones accessories internal batteries}''. Finally, for the convenience of modeling the sequential pattern, we randomly selected one as the final query for the products with multiple queries.

\subsubsection{Baseline Methods}
We compared the proposed ALSTP model with a logistic regression based method and different retrieval approaches from two categories: 1) traditional methods based on bag-of-words representations, such as  \emph{Query Likelihood Model}~\cite{zhai2004study} and \emph{Extended Query Likelihood with User Models}~\cite{ai2017learning}; and 2) representation learning approaches based on latent space modeling, such as \emph{Latent Semantic Entity}~\cite{van2016learning} and \emph{Hierarchical Embedding Model (HEM)}~\cite{ai2017learning}. It is worth noting that the recently proposed HEM is the state-of-the-art method for personalized product search. We introduced these models in the following. To ensure a fair comparison, we carefully tuned these models and reported the best performance.

\textbf{Logistic Regression Search Model (LRS).} This method is the variant of the winner of CIKM Cup 2016 competition. The original method is designed for click prediction tasks, we adapted it to the same setting with our proposed method as well as other baselines. Due to the lack of abundant auxiliary information, we adopted the second best model in the original paper~\cite{wu2017ensemble}, of which the final performance is 0.4175 of NDCG (The best model is 0.4238) on the competition dataset\footnote{\href{https://competitions.codalab.org/competitions/11161}{https://competitions.codalab.org/competitions/11161.}}. The query representations obtained via the PV-DM model are used to represent the query feature, and the product search task is converted to a classification problem. Impressive performance was observed (Details can be found at Tab~\ref{tab:baseline}).

\textbf{Query Likelihood Model (QL).} This method is a language modeling approach.  It first estimates a language model for each document, and then ranks the documents by the likelihood of generating the query according to the estimated model~\cite{zhai2004study}. Formally, given a query $\mathit{Q}$, the retrieval score of a document $\mathit{D}$  is defined as,
\begin{equation}
P_{QL}(Q|D) = \sum_{w \in Q}log\frac{tf_{w, D} + \mu{P(w|C)}}{|D| + \mu},
\end{equation}
where $tf_{w, D}$ is the frequency of the word $w$ in $D$, $|D|$ is the length of $D$, $\mu$ is a Dirichlet smoothing parameter, and $P(w|C)$ is a collection language model which can be computed as the frequency of $w$ in the corpus $C$ divided by the size of $C$. The optimal value of $\mu$ varies from collection to collection. In our experiment, we tested different values from 2,000 to 10,000 with a step size of 4,000.

\textbf{Extended Query Likelihood with User Models (UQL).} This model is first introduced to the personalized product search by Ai et al.~\cite{ai2017learning}. Specifically, let $U$ be the set of the most frequent words\footnote{Here, we define the words appearing more than 50 times as the most frequent words.} of reviews submitted by the user $u$, and then the likelihood model of the user-query pair $(U, Q)$ for document $D$ is
\begin{equation}\label{}
P_{UQL} = \lambda{P}_{QL}(Q|D) + (1-\lambda)P_{QL}(U|D),
\end{equation}
where $\lambda$ is a coefficient parameter controlling the weights of $U$ in the search. We searched $\lambda$ in $[0, 1]$ with a step size of 0.2. When $\lambda=1$, it becomes the same as the $QL$ method.

\textbf{Latent Semantic Entity (LSE).} This method is specially designed for product search~\cite{van2016learning}. It maps  words and products into the same latent space and learns a mapping function $f_{LSE}$ between them by,
\begin{equation}\label{}
f_{LSE}(s) = tanh(\bm{W}(\bm{W_{v}}\frac{1}{|s|}\sum_{w_{i} \in s}\bm{\delta_{i}}) + \bm{b}),
\end{equation}
where $s$ is a $n$-gram string extracted from the review of an item or a user-issued query, $w_i$ is the $i$-th constituent word in $s$, and $f_{LSE}$ is the learned representation of $s$ in the latent space. The objective is to directly maximize the similarity between the vector representation $\bm{e_i}$ of the product $i$ and its corresponding projected $n$-gram $s_i$'s latent representation $f_{LSE}(s_i)$. For simplicity, we set the word embedding and product embedding to the same size; and for the $n$-gram window size $n$, we tuned it exponentially $\{2^i$ | $2 \leq i\leq 4\}$.

\textbf{Hierarchical Embedding Model (HEM).} This model (HEM) proposed by~\cite{ai2017learning} is the state-of-the-art approach for the personalized product search. It extends LSE~\cite{van2016learning} by adding the element of user preference to the product search. Similar to UQL, HEM also uses a coefficient to control the weight between the query model $\bm{q}$ and the user model $\bm{u}$ by,
\begin{equation}\label{}
\bm{M}_{\bm{uq}} = \lambda\bm{q} + (1 - \lambda)\bm{u}.
\end{equation}
HEM learns the distributed representations of queries, users and products by maximizing the likelihood of observed user-query-product triplets. We tuned the query model weight $\lambda$ from $0.0$ to $1.0$ with a step size of $0.2$.

\subsubsection{Evaluation Metrics}
We applied three standard metrics to measure the performance of our model and the baselines from distinct perspectives: Hit Ratio (HR), Mean Reciprocal Rank (MRR), and Normalized Discounted Cumulative Gain (NDCG).
\begin{itemize} [align=left,style=nextline,leftmargin=*,labelsep=\parindent,font=\normalfont]
	\item \textbf{HR} measures whether the top 20 results contain the correct items. In our setting, it indicates the percentage of queries to which the method hits a correct item in the list.
	\item \textbf{MRR} is a popular metric in the information retrieval field. It is the average of reciprocal ranks of the desired products.
	\item \textbf{NDCG} is widely used for measuring the rank accuracy, as it takes into account the position of positive items in the rank list by assigning a higher score to the item at a higher position.
\end{itemize}
Without special specification, we truncated the rank list at 20 for all the three metrics.

\subsubsection{Experimental Settings}
\textbf{Dataset Split.}
We partitioned each of the four datasets into three sets: training, validation and testing sets. We first extracted the user-product pairs from users' reviews, and then extracted the queries for these products. We finally got the user-query-product triplets. For each dataset, the last purchasing transaction of each user is held for the testing set, the second last for the validation set, and the rest for the training set. Moreover, we hid the reviews of the validation and testing sets in the training phase to simulate the real-world scenarios. For the ALSTP model, we trained it on the training set, tuned the parameters on the validation set, and reported the final results on the testing set based on the optimal parameter settings.

\textbf{Parameter Settings.} In the training procedure of ALSTP, the parameters are initialized by the $xavier$ method~\cite{glorot2010understanding} and then optimized with the standard Stochastic Gradient Descent (SGD) with the momentum value 0.9. The layers of GRU is fixed to 1, the learning rate is tuned in the range of [0.00001, 0.0005, 0.0001, 0.001, 0.01], and the regularizer is [0.000001, 0.00001, 0.0001, 0.001].  The number of negative samples (i.e., $N_s$) for each positive training data is set to 5 for LSE, HEM and our model. Besides, to avoid the unstable gradient update, we clipped the global norm of parameter gradients with 5~\cite{pascanu2013difficulty}.
\begin{table}
	\caption{Performance comparisons between ALSTP and baselines over four Amazon datasets.  Symbols $\ast$ and $\dag$ denote the statistical significance with two-sided t-test of $\mathit{p} < 0.05$ and $\mathit{p} < 0.01$, respectively, compared to the best baseline. The best performance is highlighted in boldface.}
	\label{tab:baseline}
	\begin{tabular}{r|r|c|c|c|c|c|c}
		
		\hline
		Dataset & Metric & LRS & QL & UQL & LSE & HEM & ALSTP \\
		\hline
		\multirow{3}{*}{Phones}& HR   & 0.050 & 0.058 & 0.058 & 0.094 & 0.176 & \textbf{0.194}$^\ast$\\
		                       & MRR  & 0.011 & 0.015 & 0.015 & 0.025 & 0.056 & \textbf{0.065}$^\dag$\\
		                       & NDCG & 0.022 & 0.024 & 0.024 & 0.040 & 0.083 & \textbf{0.096}$^\dag$\\
		\hline
		\multirow{3}{*}{Clothing} & HR  & 0.076 & 0.079 & 0.082 & 0.040 & 0.076 & \textbf{0.109}$^\dag$\\
		                          & MRR & 0.015 & 0.019 & 0.019 & 0.012 & 0.014 &\textbf{0.037}\\
		                          & NDCG& 0.027 & 0.032 & 0.033 & 0.023 & 0.027 & \textbf{0.052}$^\dag$\\
		\hline
		\multirow{3}{*}{Toys} & HR  & 0.150 & 0.105 & 0.112 & 0.065 & 0.188 & \textbf{0.202}$^\dag$\\
		                      & MRR & 0.042 & 0.030 & 0.031 & 0.013 & 0.050 & \textbf{0.061}$^\dag$\\
		                      & NDCG& 0.076 & 0.047 & 0.049 & 0.024 & 0.084 & \textbf{0.094}$^\dag$\\
		\hline
		\multirow{3}{*}{Electronics} & HR  & 0.069 & 0.080 & 0.080 & 0.110 & 0.179 & \textbf{0.198}$^\dag$ \\
		                             & MRR & 0.016 & 0.021 & 0.021 & 0.023 & 0.055 & \textbf{0.064}$^\ast$\\
		                             & NDCG& 0.028 & 0.034 & 0.034 & 0.042 & 0.083 & \textbf{0.091}$^\dag$\\
		\hline
	\end{tabular}
\end{table} 

\subsection{Performance Comparison (RQ1)} \label{sect:comp}
Table~\ref{tab:baseline} summarizes the comparison results between our model and all the baselines over the four datasets regarding all the three metrics. We also conducted pairwise significance test (i.e., t-test) between our model and the baseline with the best performance. The main observations from Table~\ref{tab:baseline} are as follows:
\begin{itemize} [align=left,style=nextline,leftmargin=*,labelsep=\parindent,font=\normalfont]
\item Across all the four datasets, our method can outperform all the competitors significantly. This reveals the integration of the current query with the attentive short and long-term user preferences can better express the user's shopping need. Compared to HEM, our method can achieve a larger improvement on the \emph{Clothing} dataset (the absolute improvement of NDCG over \emph{Clothing} is 0.025 (93\% relative improvement) while on \emph{Phone} is 0.013 (16\%), \emph{Toys} is 0.010 (12\%), and \emph{Electronics} is 0.008 (10\%)), which is mainly because of the more frequent behaviors of purchasing clothes. Besides, \emph{fashion} and \emph{seasonal changes} exert more influence on user's local choices on clothes, revealing the importance of the short-term user preference.
\item For both the traditional bag-of-words (i.e., QL and UQL) and state-of-the-art representation learning (LSE and HEM) methods, personalized product search consistently outperforms the non-personalized ones. This indicates that the user's shopping intention regarding the same query can be much diverse as well as the importance of personalization in the product search.
\item For the logistic regression based method LRS, on the \emph{Toys} dataset, it surpasses the QL and UQL with a large margin, and achieved better performance than LSE and HEM on the \emph{Clothing} dataset. This is mainly because the number of products per query on these two datasets is much shorter than the other two datasets. So that the concise logistic regression method can generalize well on these two datasets.
\item On the \emph{Clothing} dataset, the performance of both  bag-of-words methods (QL and UQL) exceed  the two representation learning methods (LSE and HEM), which is mainly because the average review length is much smaller than that of the other three datasets (\emph{Clothing} is 36, while on the other three are 135, 103, 161, respectively). This indicates that less review words make the user preference modeling more difficult for the representation learning methods. Our model not only leverages the review word embedding to learn the product representation, but also considers the recently purchased products to have more insights the user's purchasing intention. Therefore, our ALSTP model can consistently outperform the QL and UQL.
\end{itemize}

\subsection{Model Ablation (RQ2 \& RQ3)} \label{sect:ablation}
To study the utility of each component in our model, we decomposed our proposed ALSTP model with a set of variants, including:
\begin{itemize} [align=left,style=nextline,leftmargin=*,labelsep=\parindent,font=\normalfont]
	\item \textbf{Without Preference Modeling (WoPM):} It prunes the user preferences (including the long-term and short-term ones) from ALSTP;
	\item \textbf{Short-Term Preference Modeling (STPM):} It is a variant of ALSTP without considering the long-term user preference or the attention mechanism. In addition, it leverages the recently bought products to denote the short-term user preference;
    \item \textbf{Attentive Short-Term Preference Modeling (ASTP):} It is a variant of ALSTP without considering the long-term user preference but the attention mechanism for the short-term preference;
	\item \textbf{Long-Term Preference Modeling (LTPM):} It is a variant of ALSTP without considering the short-term user preference or the attention mechanism. It leverages the user's whole transaction history to learn the long-term user preference;
	\item \textbf{Attentive Long-Term Preference Modeling (ALTP):} It is a variant of ALSTP without considering the short-term user preference but the attention mechanism for the long-term user preference;
	\item \textbf{Long Short-Term Preference Modeling (LSTP):} It prunes the attention mechanism from our proposed ALSTP model.
\end{itemize}

\begin{figure*}
	\centering
	\includegraphics[width=0.9\textwidth,keepaspectratio]{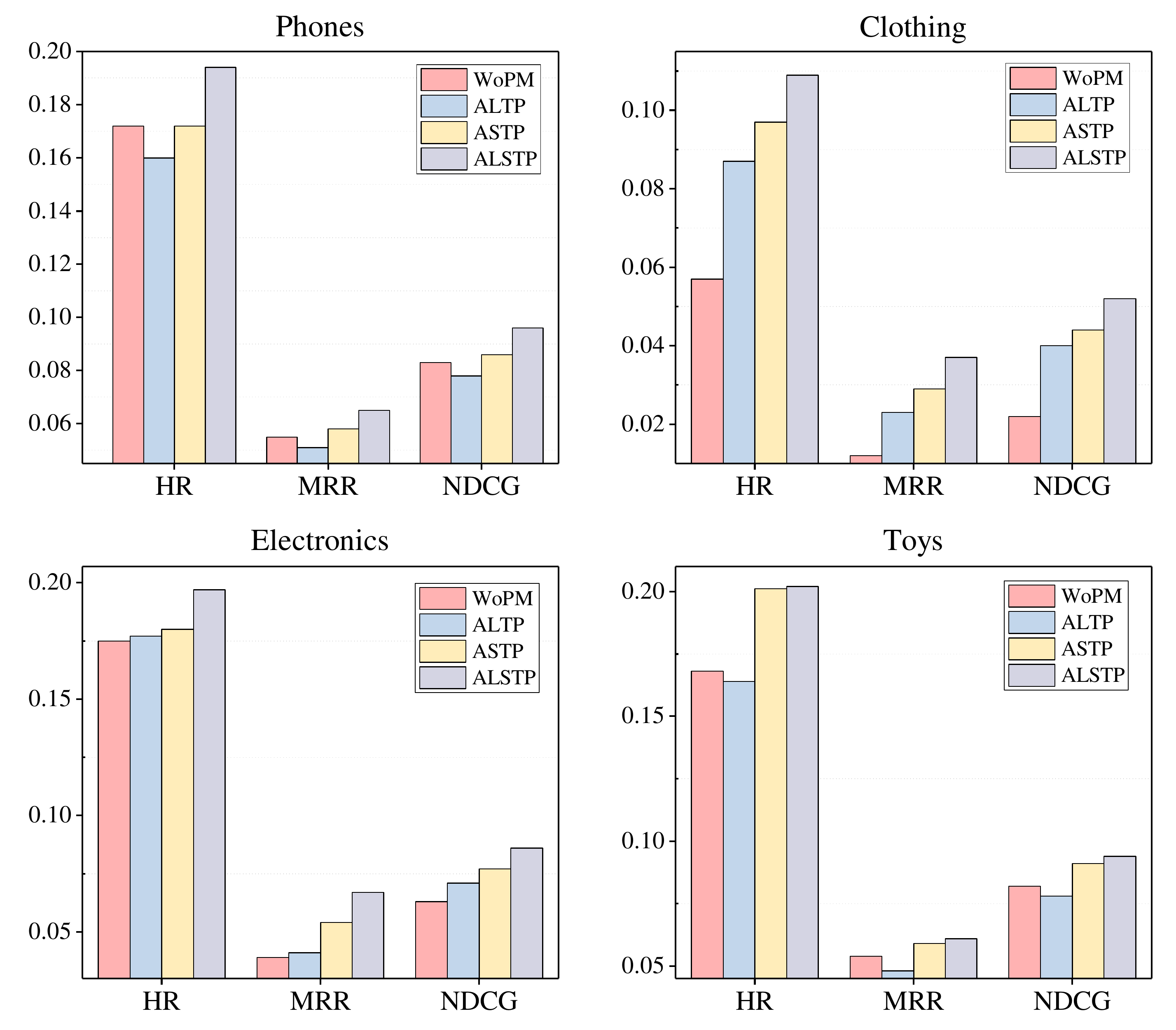}
	\caption{Comparison of variants of our model.}
    \label{fig:bar}
\end{figure*}
\textbf{Comparison Among Preferences.} We compared the ALSTP with three variants: WoPM, ASTP, ALTP, representing product search without preferences, with short-term user preference, and with long-term user preference, respectively. The performance of these three variants and our final model is shown in Figure~\ref{fig:bar}. The main observations are three-fold:
\begin{itemize} [align=left,style=nextline,leftmargin=*,labelsep=\parindent,font=\normalfont]
    \item With the consideration of the user's shopping interest in a short term, i.e. ASTP, we observed it has better performance than WoPM and ALTP on the four datasets. It reveals the influence of user's previous purchases on the next purchasing intention, and the importance of considering user's short-term preference in the product search.
    \item On datasets of \emph{Phones} and \emph{Toys}, ALTP (with the consideration of the long-term preferences) performs worse than WoPM. This is mainly because the incidently transient events (which can be modeled through the short-term user preference) affect user's decisions more and the long-term user preference has not updated promptly. Therefore, the long-term user preference would disturb the current query and may cause the drift of the query intention.
    \item For WoPM, though it simply matches the learned representations of queries and products without personalization, it can still perform well comparing to the baselines in Table~\ref{tab:baseline}. This may be credited to the power of the representation learning of our neural-net framework.
\end{itemize}

\textbf{Utility of Attention Mechanism.} We analyzed the utility of the attention mechanism for ASTPM, ALTPM and ALSTP, corresponding to the short-term user preference modeling, long-term preference user modeling and our final model, respectively. As we can see from Table~\ref{tab:attention}, the models with attention perform better than those without attention in most cases.  As we mentioned, the long-term preference contains diverse manifestations. Without the specific emphasis on relevant aspects in the long-term user preference, it may also lead to the ``query drift'' problem. For the short-term user preference, the previously bought products do not contribute equally to the next purchase. Hence, it is also important to treat them differently regarding the current query. For the proposed ALSTP model, the attention mechanism is necessary. It helps us capture not only the previous important local factors in the short-term user preference, but also the triggering aspects in the long-term user preference. For example, on the \emph{Electronics} dataset, the relative improvement of ALSTP over LSTP is 23\% for MRR, ASTP over STPM is 4\%, and ALTP over LTPM is 7\%.

\begin{table}
	\caption{Influence of attention mechanism. Symbols $\ast$ and $\dag$ denote the statistical significance with two-sided t-test of $\mathit{p} < 0.05$ and $\mathit{p} < 0.01$, respectively, compared to the methods without attention mechanism.}
	\label{tab:attention}
	\begin{tabular}{r|r|c|c|c|c|c|c}
		\hline
		\multirow{2}{*}{Dataset}  & \multirow{2}{*}{Metric} & \multicolumn{2}{c|}{ASTPM} &
                                    \multicolumn{2}{c|}{ALTPM} & \multicolumn{2}{c}{ALSTP}\\
                                    \cline{3-8}
        && STPM & ASTP & LTPM & ALTP & LSTP & ALSTP \\ \cline{3-4} \cline{5-6} \cline{7-8}
		\hline
		\multirow{3}{*}{Phones}       & HR  & 0.173 & 0.173$^\ast$ & 0.158 & 0.160$^\ast$ & 0.183 & 0.194$^\ast$ \\
		                              & MRR & 0.056 & 0.058 & 0.048 & 0.051$^\dag$ & 0.061 & 0.065$^\dag$ \\
		                              & NDCG& 0.085 & 0.086$^\ast$ & 0.075 & 0.073$^\dag$ & 0.090 & 0.096$^\dag$ \\
		\hline
		\multirow{3}{*}{Clothing}     & HR  & 0.092 & 0.097$^\ast$ & 0.083 & 0.087$^\dag$ & 0.100 & 0.109$^\ast$ \\
		                              & MRR & 0.029 & 0.027$^\ast$ & 0.021 & 0.023$^\dag$ & 0.033 & 0.037$^\dag$ \\
		                              & NDCG& 0.042 & 0.044$^\ast$ & 0.034 & 0.040 & 0.047 & 0.052 \\
		\hline
		\multirow{3}{*}{Toys}         & HR  & 0.192 & 0.201$^\ast$ & 0.159 & 0.164$^\dag$ & 0.200 & 0.202$^\dag$ \\
		                              & MRR & 0.059 & 0.061$^\ast$ & 0.045 & 0.048$^\ast$ & 0.059 & 0.061$^\dag$ \\
		                              & NDCG& 0.091 & 0.094 & 0.072 & 0.078$^\dag$ & 0.092 & 0.094 \\
		\hline
		\multirow{3}{*}{Electronics}  & HR  & 0.173 & 0.180$^\dag$ & 0.174 & 0.178$^\dag$ & 0.190 & 0.198$^\dag$ \\
		                              & MRR & 0.054 & 0.056$^\dag$ & 0.042 & 0.045$^\ast$ & 0.052 & 0.064$^\dag$ \\
		                              & NDCG& 0.080 & 0.083$^\dag$ & 0.068 & 0.070$^\ast$ & 0.079 & 0.091$^\dag$ \\
		\hline
	\end{tabular}
\end{table}

\begin{figure*}
	\centering
	\includegraphics[width=0.9\textwidth,keepaspectratio]{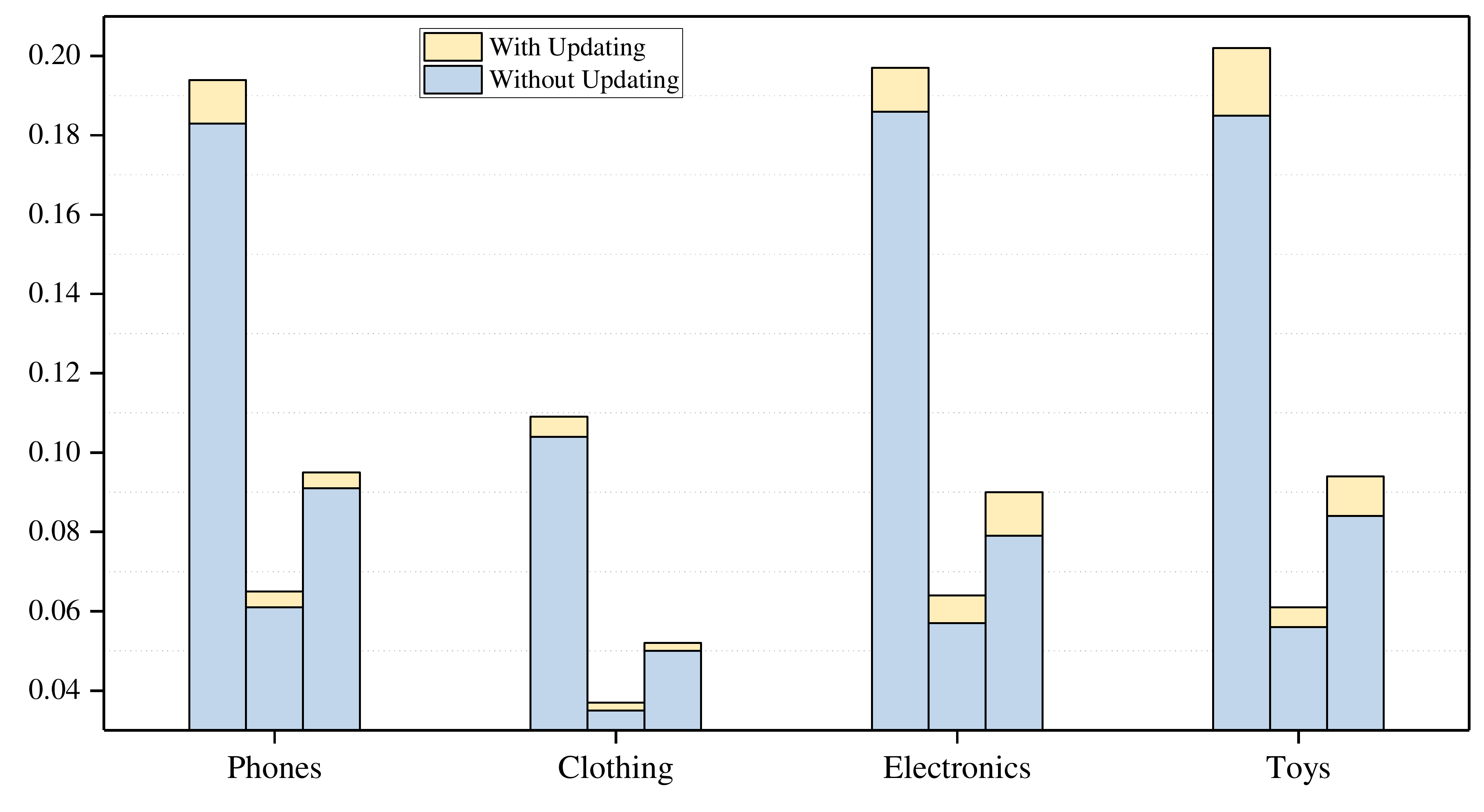}
	\caption{Influence of the long-term user preference updating.}
    \label{fig:update}
\end{figure*}

\textbf{Utility of Long-term User Preference Updating.} We studied the utility of the long-term user preference updating on the four datasets with respect to the three metrics by tuning the updating rate $\beta$~\footnote{Although the consideration of updating rate can obtain better performance, there is no uniform trends observed with the changing of $\beta$. Besides, the optimal value of $\beta$ are different from dataset to dataset. For simplicity, we reported the results based on a fixed value of $\beta$ (i.e., $\beta$ =0.5), ), based on which a relatively good performance can be observed for all the four datasets.} and reported the performance with the long-term user preference updating and the performance without updating. As shown in Figure~\ref{fig:update}, the performance with the long-term user preference updating surpasses the one without updating. This is in accordance with our assumption that the long-term user preference should update gradually.

\subsection{Influence of Important Parameters (RQ4)} \label{sect:parameters}
\begin{figure*}
	\centering
	\includegraphics[width=1.0\textwidth,keepaspectratio]{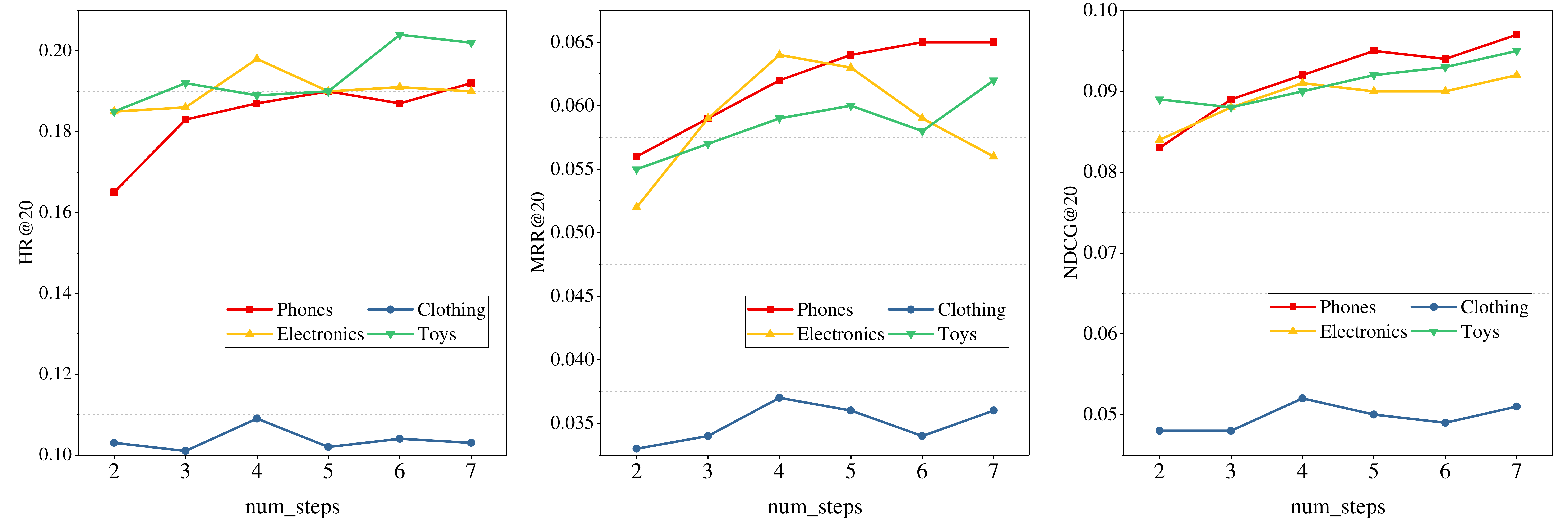}
	\caption{Performance of ALSTP on four Amazon datasets by varying the numbers of previous queries with respect to
 HR, MRR and NDCG.}
    \label{fig:steps}
\end{figure*}

\begin{figure*}
	\centering
	\includegraphics[width=0.95\textwidth,keepaspectratio]{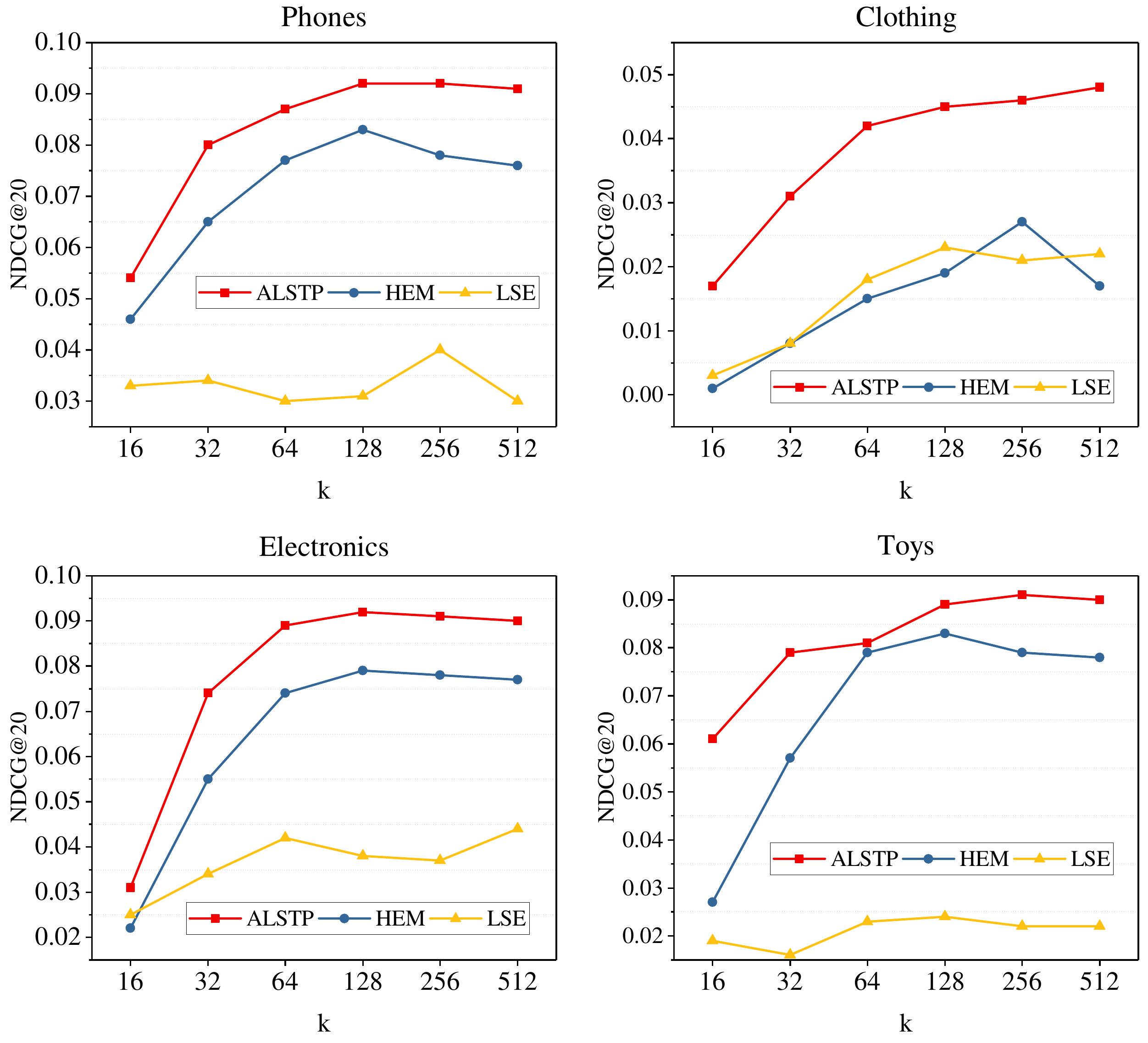}
	\caption{Performance of LSE, HEM and ALSTP on four Amazon datasets by varying the embedding size of user preference with respect to NDCG.}
    \label{fig:dimension}
\end{figure*}
In this section, we studied the influence of two parts: 1) \emph{the number of previous purchase queries/items in short-term preference modeling}; and 2) \emph{the embedding size of the user preference on the final performance of our model}.

\textbf{Number of Previous Queries.}
As show in Figure~\ref{fig:steps}, the effect of the number of previous queries varies considerably on different datasets. For the datasets like \emph{Phones} and \emph{Toys}, user's purchasing behaviors are relatively sparse. For example, a user will not buy phones, phone accessories, or dolls frequently in a short time.  Therefore, considering more previous queries could provide more information to infer the user's current preference (i.e., \emph{Phones} for seven and \emph{Kindle} for six). While for the other two datasets, there are more short-term patterns in the user's purchase behaviors. Taking \emph{Clothing} for example, users tend to buy a set of clothes in a short period, showing the obvious local preference. Due to a forthcoming event (e.g., \emph{wedding} or \emph{new season}), they will change to purchase a new style of clothes.

\textbf{Embedding Size of User Preference.} To analyze the effect of the embedding size on the baselines of LSE and HEM, as well as our proposed ALSTP model, we show the results of these methods with distinct embedding sizes over the four datasets. Figure~\ref{fig:dimension} shows the performance variation with the increasing of the embedding sizes. It can be observed that for all the three methods, with the increasing of the embedding size, the performance improves firstly, and then starts to deteriorate. Generally, the larger embedding size will lead to the better representation capability, while it will result in over-fitting when the embedding size is too large. From the Figure~\ref{fig:dimension}, we can see that 256 is a proper embedding size for our model.

\begin{figure}
	\includegraphics[width=0.9\textwidth]{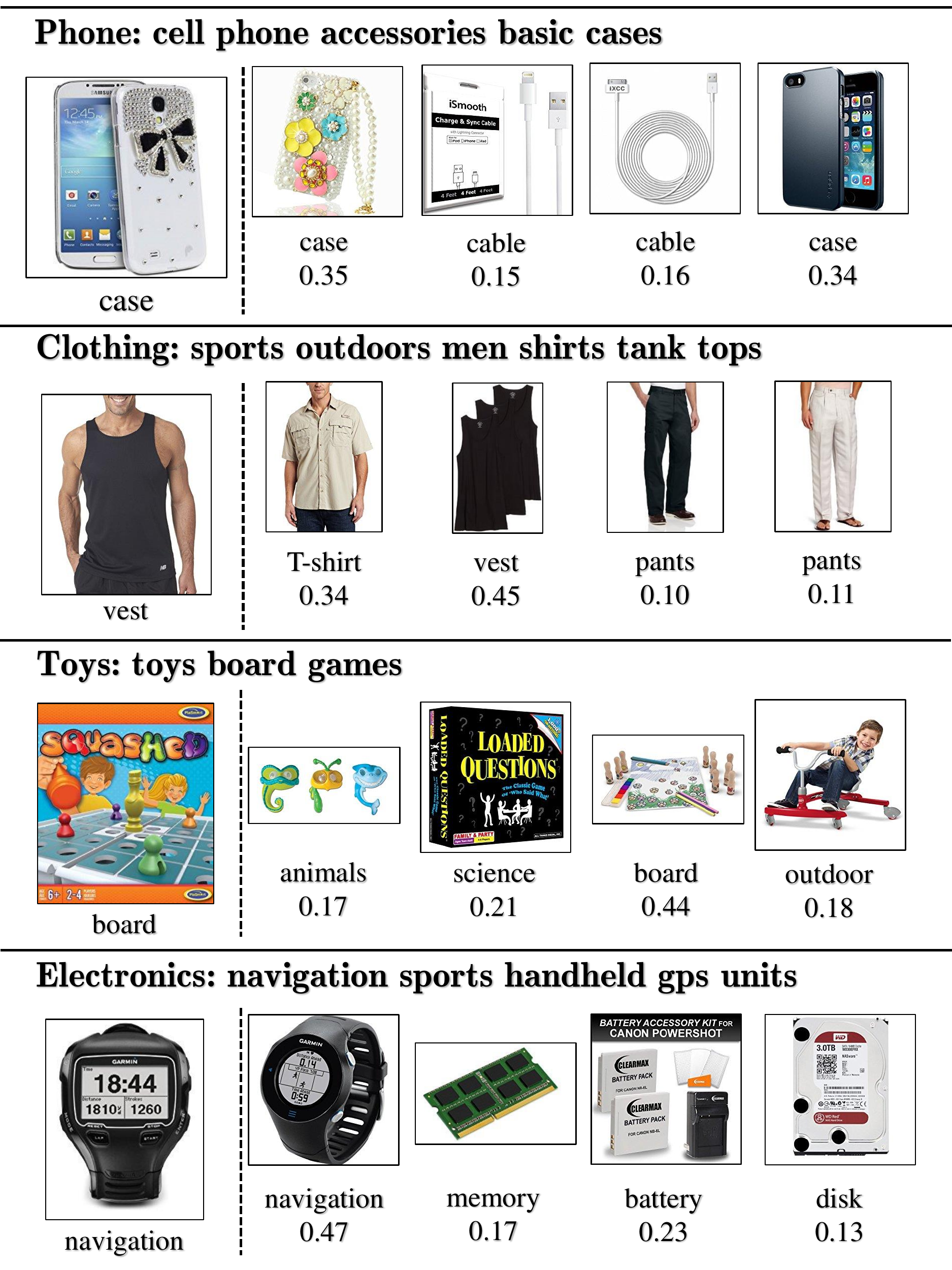}
	\caption{Visualization results of the ASTPM model over four datasets. Here, the first line shows the chosen queries of each dataset, and we removed some words (i.e., words of category information) in the query for simplicity. The picture on the left side is the current product that the user would like to buy; the products on the right are immediately previously purchased products; values in the last line are the corresponding attention weights.}
	\label{fig:viz}
\end{figure}

\subsection{Short-term Preference Attention Visualization (RQ5)} \label{sect:examples}
To demonstrate the effectiveness of our attention mechanism on the short-term user preference, we illustrated a few examples. We randomly chose one example from each dataset. As shown in Figure~\ref{fig:viz}, although our attention weights are computed based on the extracted queries (corresponding to each product), we could still observe the relevance between the desired product and the previously purchased products. For the first query, the user wants to search \emph{cell phone basic cases}, based on his/her former purchased \emph{phone cases} and \emph{phone cable}, we can see the two previously purchased \emph{phone case}s are more related to the desired product than the others. In the second example, for the currently targeted search product \emph{men's vest}, the previously purchased product like \emph{men's long pants} and \emph{men's T-shirt} are less related, while the recently bought \emph{vest} is more useful on suggesting the targeted product. The user in the third example shows diverse interests in \emph{toys \& games}, and intends to find \emph{board games}. Obviously, the third item is more related compared to the other three ones. In the last example, the user intends to buy a \emph{navigation sports handheld} instead of \emph{computer memory}, \emph{battery} or \emph{hard disk}, thus the purchased \emph{navigation} is more informative in guiding the current choice.

Based upon the above observations and analysis, our attentive short-term preference modeling module is verified to be quite useful on inferring user's current search intention.

\section{Conclusion and future work}\label{conclusion}
In this paper, we present an ALSTP model to better capture the user preference for personalized product search. We argue that it is critical to consider the user's long- and short-term preferences simultaneously in the product search. To this end, we introduced a novel ALSTP model to capture the attentions of the long- and short-term user preferences on the current query, and then attentively integrated them with the current query to better represent the user's search intention. To the best of our knowledge, ALSTP is the first model that exploits the attention mechanism on both user's short and long-term preferences for personalized product search. We conducted extensive experiments on four real-world Amazon Datasets and demonstrated that ALSTP can remarkably outperform several state-of-the-art competitors in the product search. Besides, we analyzed the effectiveness of each component in our model and visualized the attentions in the short-term preference modeling. In the future, we plan to leverage a pre-trained $RNN$ to better model the short-term user preference. Moreover, encoding user's demographic information into the long-term user preference modeling and semantic understanding of recently purchased products into the short-term user preference modeling may lead to the better performance.

\begin{acks}
This work is supported by the National Basic Research Program of China (973 Program), No.: 2015CB352502; National Natural Science Foundation of China, No.: 61772310, No.:61702300, No.:61672322, and No.:61702302; the Project of Thousand Youth Talents 2016; the Tencent AI Lab Rhino-Bird Joint Research Program, No.:JR201805; and the National Research Foundation, Prime Minister’s Office, Singapore under its International Research Centre in Singapore Funding Initiative.
\end{acks}

\bibliographystyle{ACM-Reference-Format}
\bibliography{ALSTP_bib}
\end{document}